\documentclass[letterpaper,11pt,twoside]{article}
 
\usepackage{amsmath,amsthm,amssymb}
\usepackage{graphicx}
\usepackage{bbm}
\usepackage{hyperref}
\usepackage{color}
\usepackage[utf8]{inputenc}
\usepackage{verbatim}
\newcommand{\doubleplus}{{+\!\!\!+}}
\newcommand{\p}{\partial}
\newcommand{\rx}{r}
\newcommand{\trx}{{\tilde r}}
\newcommand{\tell}{{\tilde\ell}}

\newcommand{\tr}{\operatorname{tr}}

\newcommand{\re}{\operatorname{Re}}
\newcommand{\im}{\operatorname{Im}}

\newcommand\F {{\cal F}}
\newcommand\G {{\cal G}}
\renewcommand\L {{\cal L}}

\newcommand\N {{\cal N}}

\newcommand\fg {{\mathfrak g}}

\newcommand\bC {{\mathbb C}}
\newcommand\bD {{\mathbb D}}
\newcommand\bF {{\mathbb F}}
\newcommand\bG {{\mathbb G}}
\newcommand\bJ {{\mathbb J}}

\newcommand\bV {{\mathbb V}}

\newcommand\nn {\nonumber}
\newcommand\ds {\displaystyle}

\let\svthefootnote\thefootnote
\newcommand\blankfootnote[1]{%
  \let\thefootnote\relax\footnotetext{#1}%
  \let\thefootnote\svthefootnote%
}

\begin{document}
\renewcommand{\theequation}{\thesection.\arabic{equation}}
\setcounter{page}{0}
\thispagestyle{empty}
%\maketitle
\begin{flushright} \small
YITP-SB-18-06\\
%Draft dated \today
\end{flushright}
\vskip15mm
\begin{center}
{\huge\bf Generalized K\"ahler structures on group manifolds and T-duality} \\
\vskip10mm
{\Large J.~P. Ang${}^*$, Sibylle Driezen${}^{\dag,\diamond}$, 
Martin Ro\v cek${}^*$, Alexander Sevrin${}^{\dag,\ddagger}$} \\
\vskip15mm
{${}^*$C.~N.~Yang Institute for Theoretical Physics, Stony Brook University \\ Stony Brook, NY 11794, USA} \\
\vskip5mm
{${}^{\dag}$Theoretische Natuurkunde, Vrije Universiteit Brussel and The International Solvay Institutes \\ Pleinlaan 2, B-1050 Brussels, Belgium} \\
\vskip5mm
{${}^{\diamond}$Department of Physics, Swansea University\\ Singleton Park, Swansea SA2 8PP, UK}
\vskip5mm
{${}^{\ddagger}$Physics Department, Universiteit Antwerpen, Campus Groenenborger \\ 2020 Antwerpen, Belgium}
\vskip10mm
\end{center}

\blankfootnote{
\begin{tabular}{ll}
Email: & \texttt{\href{mailto:JianPeng.Ang@gmail.com}{JianPeng.Ang@gmail.com}, \href{mailto:Sibylle.Driezen@vub.be}{Sibylle.Driezen@vub.be}}, \\
& \texttt{\href{mailto:Martin.Rocek@stonybrook.edu}{Martin.Rocek@stonybrook.edu}, \href{mailto:Alexandre.Sevrin@vub.be}{Alexandre.Sevrin@vub.be}}
\end{tabular}}
\date{\today}
\vskip1cm
\begin{center}
{\large\bf Abstract} \\[5mm]
\end{center}
We study generalized K\"ahler structures on $N=(2,2)$ supersymmetric Wess-Zumino-Witten models; we use the well known case of $SU(2)\times U(1)$ as a toy model and develop tools that allow us to construct the superspace action and uncover the highly nontrivial structure of the hitherto unexplored case of $SU(3)$; these tools should be useful for studying many other examples. 
We find that {\em different} generalized K\"ahler structures on $N=(2,2)$ supersymmetric Wess-Zumino-Witten models can be found by T-duality transformations along affine isometries. 
\newpage
\setcounter{tocdepth}{2}
\tableofcontents

\section{Introduction}
\setcounter{equation}{0}
\label{sec.intro}
Almost since its inception it was recognized that supersymmetry and geometry go hand in hand~\cite{Zumino:1979et}. Supersymmetric non-linear $\sigma$-models in two dimensions (NLSM) are a class of field theories where the geometric aspects are under control and can be studied both at the classical and the quantum mechanical level. They are interesting in their own right, and have many applications, including as the building blocks for type II string theories, as the description of certain moduli spaces,  condensed matter physics, {\it etc.}

A non-supersymmetric NLSM is fully characterized by its target manifold, which is endowed with a metric and a closed 3-form. These models can always be supersymmetrized as long as the number of supersymmetries is bounded by $(N_+,N_-)\leq (1,1)$ where $N_+$ ($N_-$) is the number of right-handed (left-handed) supersymmetries. No further geometric structure arises at the classical level. However, any additional supersymmetry past the first introduces a covariantly constant complex structure,\footnote{In general, covariantly constant with respect to a connection with torsion related to the closed 3-form.} with respect to which the metric is hermitian. In addition, if $(N_+,N_-)\geq (2,2)$, there are complex structures of each handedness. 

In the current paper we will mostly focus on the $(N_+,N_-)=(2,2)$ case, which requires besides the metric and the closed 3-form, two covariantly constant complex structures that both preserve metric, hence the name ``bihermitian geometry''. A simple dimensional argument shows that the Lagrange density in $(2,2)$ superspace can only be a function of a number of (constrained) scalar superfields. The Lagrange density encodes the full local geometry. In the simplest case where only chiral superfields appear, the two complex structures coincide, the 3-form vanishes and the geometry is K\"ahler. The Lagrange density is then precisely the K\"ahler potential. This suggests that the generic case describes a far reaching generalization of K\"ahler manifolds where the Lagrange density gets the interpretation of a generalized K\"ahler potential. 
This was understood to be the case in the bihermitian language of \cite{Gates:1984nk} in a series of papers from the NLSM perspective \cite{Buscher:1987uw}\cite{Sevrin:1996jr}\cite{Grisaru:1997ep}\cite{Lindstrom:2005zr}. This was reinterpreted by Hitchin when he introduced the concept of generalized K\"ahler geometry \cite{Hitchin:2004ut}, a natural generalization of K\"ahler geometry acting non-trivially on the sum of the tangent and cotangent bundle, and which was shown by Gualtieri \cite{Gualtieri:2003dx} to coincide with bihermitian geometry. A subclass of generalized K\"ahler manifolds, generalized Calabi-Yau manifolds, are conformally invariant at the quantum level and provide an important class of supergravity solutions \cite{Grana:2004bg,Jeschek:2004wy}.

The off-shell completion of a $(2,2)$ NLSM depends on the precise choice of the right and left complex structures $J_+$ and $J_-$; when different choices are possible, they correspond to different bihermitian structures and thus to different generalized K\"ahler structures. This is reflected in the supersymmetry algebra: one finds it closes off-shell modulo terms proportional to the commutator of the two complex structures $[J_+,J_-]$. As a consequence one expects that $\ker[J_+,J_-]= \ker(J_+- J_-)\oplus \ker (J_++J_-)$ can be described in a manifestly supersymmetric way without introducing any further off-shell degrees of freedom. This is indeed achieved by means of chiral and twisted chiral superfields \cite{Gates:1984nk}. To close the supersymmetry algebra off-shell when the image $[J_+,J_-]$ is nonvanishing, one must introduce additional $(1,1)$ auxiliary fields;  in $(2,2)$, they arise from semi-chiral superfields \cite{Buscher:1987uw}. Any $(2,2)$ NLSM can be described in terms of these three classes of superfields: chiral, twisted chiral and semi-chiral \cite{Lindstrom:2005zr}. However, as the off-shell completion of a $(2,2)$ NLSM fully depends on the choice made for $J_+$ and $J_-$ and as this choice is not always unique, one finds that a given target manifold often admits different generalized K\"ahler structures. 

If the generalized K\"ahler manifold possesses an isometry, one can T-dualize the model along that isometry \cite{Buscher:1987qj}. Generically one ends up with a different manifold. T-duality not only affects the metric and closed 3-form, but also acts non-trivially on the complex structures \cite{Ivanov:1994ec}. Hence, T-duality alters the superfield content of a $(2,2)$ NLSM. Two cases appear: a chiral superfield can be interchanged for a twisted chiral superfield (and vice-versa) \cite{Gates:1984nk} or a pair of chiral and twisted chiral superfields gets exchanged for a semi-chiral multiplet (and vice-versa) \cite{Grisaru:1997ep}.  A particularly interesting case arises when the isometry is actually a Kac-Moody symmetry -- then the metric and closed 3-form remain unchanged, but the complex structures still transform 
\cite{Rocek:1991ps}. This is precisely the case we investigate in this paper.

A  simple but non-trivial class of generalized K\"ahler manifolds where many of the issues discussed above can be studied quite explicitly are even-dimensional reductive Lie group manifolds \cite{Spindel:1988sr}. The resulting $\sigma$-models are $(2,2)$ supersymmetric Wess-Zumino-Witten (WZW) models. A complex structure on a reductive group manifold is fully determined by its action on the Lie algebra where it is almost equivalent to a Cartan decomposition of the Lie algebra: it has eigenvalue $+i$ ($-i$) on positive (negative) roots. The only freedom remains in its action on the Cartan subalgebra where the only restriction is the requirement that the Cartan-Killing metric should be hermitian. Given a choice for $J_+$ there is still a considerable freedom in choosing $J_-$, giving rise to various generalized K\"ahler structures on reductive even-dimensional Lie groups. For groups of low rank this can be studied systematically.

In the current paper we explore and elucidate the relation between various generalized K\" ahler structures on the same Lie group. We start with the well-known example of $SU(2)\times U(1)$, which allows for two generalized K\" ahler structures: one in terms of a chiral and a twisted chiral field \cite{Rocek:1991vk} and one in terms of semi-chiral multiplet \cite{Sevrin:1996jr}. We show that the two generalized K\" ahler structures are related through T-duality transformation along an affine isometry. This can be understood as follows. Only the maximal abelian subgroup of the left and right-handed affine group acts trivially on the complex structures and thus are manifest in $(2,2)$ superspace. T-dualizing along an affine isometry does not alter the metric or the closed 3-form \cite{Rocek:1991ps} but it does alter the complex structures, mapping one generalized K\" ahler structure on $SU(2)\times U(1)$ into the other one! As a far more difficult example, we study the hitherto unexplored case of $SU(3)$;  this also has (at least) two inequivalent generalized K\" ahler structures: one in terms of two semi-chiral multiplets and one in terms of a single semi-chiral multiplet, one chiral and one twisted chiral superfield.\footnote{Actually, both $SU(2)\times U(1)$ and $SU(3)$ have $(4,4)$ supersymmetry, so in principle there are $S^2\times S^2$ generalized K\"ahler structures. We expect these to fall into two deformation classes, so that our examples should be generic, but further investigation might be worthwhile.} 

The outline of the rest of the paper is as follows. Section 2 reviews supersymmetric WZW models, which form an important class of generalized K\"ahler manifolds. Section 3 comprises a review of the generalized K\"ahler structures carried by $SU(2)\times U(1)$, in particular concentrating on relating generalized K\"ahler structures of different types via T-duality. Section 4 studies  the generalized K\"ahler structures on $SU(3)$. Generalized K\"ahler potentials for both types of generalized K\"ahler structures on $SU(3)$ are derived. Section 5 has a brief summary of our results and discusses possibilities for further research. Appendix A reviews sigma models 
and their supersymmetric extensions to $(1,1)$ superspace. Appendix B gives details of the $(2,2)$ superspace description of bihermitian geometry, including explicit formulae for the complex structures $J_\pm$ in terms of the generalized K\"ahler potential $K$. Appendix C discusses isometries, T-duality, and generalized $(2,2)$ supersymmetric vector multiplets. Appendix D attempts to give insight into the art of finding holomorphic coordinates on WZW-models. Appendix E presents some particular choices of holomorphic coordinates for $SU(2)\times U(1)$ that we found but are not discussed in section 3. Appendix F describes the T-duality transformation from type $(1,1)$ to $(0,0)$ using the Large Vector Multiplet discussed in Appendix C.  Appendix G discusses another rank 2 group, $SU(2)\times SU(2)$, which admits only type $(1,0)$ and type $(0,1)$ generalized K\"ahler structures.

\section{Generalized K\"ahler geometry on group manifolds}
\setcounter{equation}{0}
\subsection{$(2,2)$ sigma model description of bihermitian geometry}
\label{sec.22nlsm}
We briefly recap the $(2,2)$ superspace formulation of a two-dimensional sigma model with bihermitian target space and establish the notation used in the rest of this paper.\footnote{For background and more details about sigma models in general, see Appendix \ref{sec.appA}.}

As always, $(2,2)$ superspace has two commuting coordinates $\sigma^\doubleplus=\tau+\sigma,\,\sigma^==\tau-\sigma$ and four anticommuting coordinates $\theta^+,\bar\theta^+,\theta^-,\bar\theta^-$. There are two complex spinorial covariant derivatives ${\bD}_\pm$ which satisfy the algebra
\begin{equation}
\{{\bD}_\pm,\bar {\bD}_\pm\}=2i\p_{\stackrel{\doubleplus}{=}},~~~~~ {\bD}_\pm^2=\bar {\bD}_\pm^2=0.
\end{equation}
The three types of $(2,2)$ superfields required to describe a generic generalized K\"ahler manifold are the following:
\begin{itemize}
\item
chiral superfields $\phi,\bar\phi$ satisfying
\begin{align}
\bar {\bD}_+\phi=0, ~~~ \bar {\bD}_-\phi=0, \nn\\
{\bD}_+\bar\phi=0,~~~ {\bD}_-\bar\phi=0,
\label{eqn.ch}
\end{align}
\item
twisted chiral superfields $\chi,\bar\chi$ satisfying
\begin{align}
\bar {\bD}_+\chi=0,~~~ {\bD}_-\chi=0, \nn\\
{\bD}_+\bar\chi=0,~~~\bar {\bD}_-\bar\chi=0,
\label{eqn.twch}
\end{align}
\item
left and right semi-chiral superfields $\ell,\bar\ell,\rx,\bar\rx$ satisfying
\begin{align}
\bar {\bD}_+\ell=0, ~~~ \bar {\bD}_-r=0, \nn \\
{\bD}_+\bar\ell =0,~~~ {\bD}_-\bar r=0.
\label{eqn.semich}
\end{align}
\end{itemize}
The action in $(2,2)$ superspace has the form
\begin{equation}
I = \int d^2\sigma\, d^4\theta\, K = \int d^2x\, \bD^2\bar \bD^2K,
\end{equation}
where $K$ is a real local function, the generalized K\"ahler potential, of the superfields $\ell,\bar\ell,\rx,\bar\rx,\phi,\bar\phi,\chi,\bar\chi$. 
The potential is defined modulo generalized K\"ahler transformations
\begin{equation}
K \mapsto K + f(\ell,\phi,\chi) + \bar f(\bar\ell,\bar\phi,\bar\chi) + g(\rx,\phi,\bar\chi) + \bar g(\bar\rx,\bar\phi,\chi),
\end{equation}
which give rise to total derivatives in the component sigma model Lagrangian density.

This single function $K$ fully encodes the local geometry of the target manifold ${\cal M}$ which must be even-dimensional and has geometric structures $(g,H,J_+,J_-)$ where\footnote{See Appendix \ref{sec.gkp} for the formulas expressing these structures in terms of the generalized potential.}
\begin{itemize}
\item
$J_+$ and $J_-$ are two integrable complex structures on ${\cal M}$ compatible with the metric $g$
\begin{equation} \begin{array}{l}
J_+^2=J_-^2=-\mathbbm{1}, \\

[X,Y]+J_\pm[J_\pm X,Y]+J_\pm[X,J_\pm Y]-[J_\pm X,J_\pm Y]=0, \\
g(J_\pm X,J_\pm Y)=g(X,Y),
\end{array} \label{eqn.compstruct} \end{equation}
where $X,Y$ are arbitrary vector fields, and

\item
$H=d^c_+\omega_+=-d^c_-\omega_-$ is a closed 3-form, where $d_\pm^c$ are the $d^c$ operators with respect to $J_\pm$, and $\omega_\pm=gJ_\pm$ are the hermitian forms of the respective complex structures.
\end{itemize}
A manifold carrying such a structure is known as bihermitian \cite{Gates:1984nk}, and has been shown to be equivalent to generalized K\"ahler geometry \cite{Gualtieri:2003dx}. Note that the second condition is equivalent to the covariant constancy
\begin{equation}
\nabla^{(\pm)}J_\pm=0
\end{equation}
of the complex structures $J_\pm$ with respect to the Bismut connections $\nabla^{(\pm)}$, which are metric connections with torsions $\pm g^{-1}H$ (first introduced by Yano -- see \cite{Yano}, pp. 150-151). Explicitly,
\begin{equation}
{\Gamma^{(\pm)}}^\mu_{\nu\rho}={\Gamma^{(0)}}^\mu_{\nu\rho}\pm \frac{1}{2}g^{\mu\sigma}H_{\sigma\nu\rho},
\end{equation}
where $\Gamma^{(0)}$ is the Levi-Civita connection. The torsion 3-form $H$ enters the sigma model description via a local 2-form potential $b$ known as the Kalb-Ramond field, with $H=db$.

A bihermitian manifold is equipped with three Poisson structures
\begin{equation}
\pi_\pm = (J_+\pm J_-)g^{-1},
\end{equation}
\begin{equation}\label{eqn.sigma}
\sigma = [J_+,J_-]g^{-1}.
\end{equation}
The superfields \eqref{eqn.ch}-\eqref{eqn.semich} arising from the $(2,2)$ superspace description may be interpreted as coordinates adapted to these Poisson structures. More specifically, near a regular point\footnote{At a regular point on a manifold with a Poisson structure, the rank of the structure is constant in a sufficiently small neighborhood of the point. Here we consider a point that is regular with respect to all three Poisson structures.}, chiral superfields are complex coordinates along $\ker\pi_-$, twisted chiral superfields are complex coordinates along $\ker\pi_+$, and semi-chiral coordinates are holomorphic Darboux coordinates along the symplectic leaves of the foliation defined by $\sigma$ \cite{Lindstrom:2005zr}. The type of the generalized K\"ahler geometry at a point is $(\dim_\bC\ker\pi_-,\dim_\bC\ker\pi_+)$; equivalently, a geometry of type $(N_c,N_t)$ at a point admits a $(2,2)$ sigma model description with $N_c$ chiral superfields and $N_t$ twisted chiral superfields near that point. In general, the type is not constant on the manifold: there may be subvarieties, known as type-change loci, on which the type increases; these must have strictly positive codimension.

\subsection{$(1,1)$ WZW models}
A Wess-Zumino-Witten (WZW) model is a theory of maps from a Riemann surface $\Sigma$ to a Lie group $G$ equipped with an invariant metric and a normalized torsion form. The $(1,1)$ WZW model is a theory of maps from a $(1,1)$ super-Riemann surface to $G$, and has action\footnote{See appendix \ref{sec.appA} for a review of $(1,1)$ superspace.}
\begin{equation}
kI[g] = -\frac{k}{\pi}\int_\Sigma d^2\sigma\, d^2\theta\, \tr (g^{-1}{\nabla}_+gg^{-1}{\nabla}_-g) - \frac{k}{\pi}\int_B d^3\tilde\sigma\, d^2\theta\, \tr(\tilde g^{-1}\p_t\tilde g\{\tilde g^{-1}\nabla_+\tilde g,\tilde g^{-1}\nabla_-\tilde g\}),
\label{eqn.11wzw}
\end{equation}
where $k$ is an integer (the level)\footnote{Nonconformal models with separate normalizations of the two terms can also be studied; their extensions to $(2,2)$ superspace are not understood.},  ``$\tr$'' is a normalized invariant bilinear form on the Lie algebra $\fg:=\operatorname{Lie}(G)$, $B$ is a 3-dimensional manifold with boundary $\p B=\Sigma$ with local coordinates $\tilde\sigma=(t,\sigma)$, and $\tilde g$ is an extension of $g$ to $B$. Modulo multiples of $2\pi i$, $I[g]$ is independent of the choice of $B$ and extension $\tilde g$.

The Maurer-Cartan forms $e_L^a,e_R^a$, defined by
\begin{equation}
g^{-1}\, dg = e_L^aT_a,~~~ dg\, g^{-1} = e_R^aT_a,
\end{equation}
where $T_a$ is a basis for the Lie algebra $\fg$, allow one to push forward tensors on the Lie algebra to the group. 

The $(1,1)$ WZW model has symmetry group $G_L\times G_R$, acting as $g\mapsto h_Lgh_R^{-1}$ (only the subgroup $(G_L\times G_R)/Z(G)$ acts nontrivially, where $Z(G)$ is the center of the group). The model also has superconformal symmetry, so the parameters $h_L$ and $h_R$ are allowed to be semilocal, satisfying
\begin{equation}
\nabla_+h_L=0,~~~\nabla_-h_R=0.
\end{equation}

\subsection{$(2,2)$ WZW models}
\label{sec.22wzw}
All even dimensional reductive Lie groups admit $(2,2)$ extensions \cite{Spindel:1988sr}. The complex structures corresponding to the extended supersymmetries \eqref{eqn.compstruct} may be pulled back to the Lie algebra
\begin{equation}
(\bJ_+)^a{}_b=(e_L)^a{}_\mu (J_+)^\mu{}_\nu (e_L^{-1})^\nu{}_b,~~~ (\bJ_-)^a{}_b=(e_R)^a{}_\mu (J_-)^\mu{}_\nu (e_R^{-1})^\nu{}_b,
\end{equation}
and in terms of the Lie algebra complex structures $\bJ_\pm$, the conditions for $(2,2)$ supersymmetry may be reformulated as \cite{Spindel:1988sr}
\begin{itemize}
\item
$\bJ_\pm$ are constant and satisfy $\bJ_\pm^2=-\mathbbm{1}$ as well as $\eta(\bJ_\pm X,\bJ_\pm Y)=\eta(X,Y)$ where $\eta$ is the Killing form and $X,Y$ are arbitrary Lie algebra elements.
\item
Further, they obey $f(X,\bJ_\pm Y,\bJ_\pm Z)+f(\bJ_\pm X,Y,\bJ_\pm Z)+f(\bJ_\pm X,\bJ_\pm Y,Z)=f(X,Y,Z)$, where $f(X,Y,Z)=\eta([X,Y],Z)$ is the alternating form constructed from the structure constants and $\eta$.
\end{itemize}
These conditions were solved in \cite{Spindel:1988sr} where it was shown that $\bJ_\pm$ may be characterized by a choice of Cartan subalgebra and positive direction. The complex structures are diagonal on the positive and negative roots with eigenvalue $+i$ and $-i$ respectively, and map the Cartan subalgebra to itself in a way that makes the Killing form hermitian. Since any two Cartan decompositions of a Lie algebra are related by group conjugation, the only freedom lies in the choice of the action on the Cartan subalgebra.

Let us now turn to the superfield content, or type, allowed for a particular WZW model. A choice of $\bJ_+$ and $\bJ_-$ on the Lie algebra fixes the superfield content. The numbers of chiral \eqref{eqn.ch} and twisted chiral superfields \eqref{eqn.twch}
\begin{equation}
N_c = \dim_\bC\ker(J_+-J_-),~~~ N_t=\dim_\bC\ker(J_++J_-)
\end{equation}
can be computed by noting that $\ker(J_+\pm J_-)=\ker(\bJ_+\pm e_Le_R^{-1}(\bJ_-) e_Re_L^{-1})$ and that $e_Le_R^{-1}$ is a transformation in the adjoint representation. The number of sets of semi-chiral superfields \eqref{eqn.semich} is then $(N-N_c-N_t)/2$, where $2N$ is the (real) dimension of the Lie group.

This can be easily analyzed for rank two groups. Here, one has essentially two choices for the Lie algebra complex structures: either they are equal $\bJ_+=\bJ_-$, or they are opposite on the Cartan subalgebra $\bJ_+|_\text{CSA}=-\bJ_-|_\text{CSA}$. For the former case, $N_t=\dim_\bC\ker(\bJ_++e_Le_R^{-1}(\bJ_-)e_Re_L^{-1})=0$ always, while $N_c=\ker(\bJ_+-e_Le_R^{-1}(\bJ_-)e_Re_L^{-1})$ can be analyzed by writing $e_Le_R^{-1}=\exp(\alpha)$ and expanding through first nontrivial order in $\alpha$. A similar analysis can be done for the latter choice. The results are given in table \ref{tbl.jrank21}. Notice that each of the rank two groups admit generalized K\"ahler structures of two different types.
\begin{table}[h]
\begin{center}
\begin{tabular}{|c|c|c|c|c|c|c|c|}
\hline 
&& \multicolumn{3}{ |c| }{$\bJ_+=\bJ_-$}& \multicolumn{3}{ |c| }{$\bJ_+\neq\bJ_-$}\\
Group& $N$ & $N_s$& $N_c$&$N_t$& $N_s$& $N_c$&$N_t$\\
\hline\hline
$SU(2)\times U(1)$&$2$&1&0&0 &0&1&1\\ 
$SU(2)\times SU(2)$&$3$&1&1&0&1&0 &1 \\ 
$SU(3)$&$4$&2&0&0 &1&1&1\\ 
$SO(5)$&$5$&2&1&0 &1&2&1\\ 
$G_2$&$7$&3&1&0&2&2&1 \\ 
\hline
\end{tabular}
\caption{The coordinate content for the rank 2 non-abelian reductive Lie groups either taking the complex structures to be equal on the Lie algebra ($\bJ_+=\bJ_-$) or having the opposite sign on the CSA ($\bJ_+\neq\bJ_-$). The number of semi-chiral, chiral and twisted chiral coordinates are denoted $N_s$, $N_c$ and $N_t$ respectively.}
\label{tbl.jrank21}
\end{center}
\end{table}

\subsection{Isometries}
\label{sec.wzwisom}
Of the $G_L\times G_R$ symmetry of the $(2,2)$ WZW model, only the subgroup $G_L\times H_R$ preserves the left complex structure $J_+$, where $H_R\subset G_R$ is the maximal torus corresponding to the action of $\bJ_+$ on the Lie algebra as described in section \ref{sec.22wzw}. A similar statement holds for $J_-$, so the group of isometries preserving both complex structures of the $(2,2)$ extended WZW is $H_L\times H_R$.\footnote{As before, the group which acts faithfully is actually $(H_L\times H_R)/Z(G)$.} Due to the superconformal invariance of the WZW model, these are in fact Kac-Moody symmetries.

\subsection{T-duality along Kac-Moody isometries}
\label{sec.wzwtdual}
Consider the T-dual sigma model of a $(2,2)$ supersymmetric WZW model along some isometry $U(1)\subset H_L\times H_R$. Since the isometry preserves the bihermitian structure, the T-duality can be performed in $(2,2)$ superspace, and the dual model also exhibits $(2,2)$ supersymmetry.\footnote{See Appendix \ref{sec.tduality} below for a review of T-duality in $(2,2)$ superspace.}

If, furthermore, the T-duality is along a left (right) Kac-Moody isometry $U(1)\subset H_L$, then in fact the metric, torsion and left (right) complex structure of the sigma model is unchanged \cite{Rocek:1991ps,Ivanov:1994ec}. Indeed, for a left Kac-Moody isometry with Killing field $k^\mu$, normalized so that it has unit norm, the chiral component of the Noether current (see (3.1) of \cite{Rocek:1991ps})
\begin{equation}
J = \left(k^\mu (g_{\mu\nu}-b_{\mu\nu})+\omega_\nu\right)\p\Phi^\nu
\end{equation}
vanishes, where $\omega$ is a one-form defined by $\L_kb=d\omega$. We assume that the $b$ field is chosen to be invariant under $k$, so that $\omega=d\alpha$ is locally exact, so
\begin{equation}
k^\mu (g_{\mu\nu}-b_{\mu\nu}) + \p_\nu\alpha = 0.
\label{eqn.chiralcurrent}
\end{equation}
Choose a coordinate system $\Phi^I$ such that $\Phi^0=-\alpha$ and the other $\Phi^i$ are $k$-invariant. Contracting \eqref{eqn.chiralcurrent} with $k^\nu$ shows that in this coordinate system, the Killing field is $k=\p/\p\Phi^0$. Taking $\nu=I$ in \eqref{eqn.chiralcurrent} then shows that the metric and $b$ field satisfy $e_{00}=g_{00}=1$ and $e_{i0}=(g-b)_{0i}=0$. Substituting $e_{00}=1$ and $e_{i0}=0$ into the formulas (12), (15), (16) of \cite{Ivanov:1994ec} then shows that the metric, torsion and left complex structure are unchanged by T-duality.

For a right Kac-Moody isometry, the antichiral component of the Noether current
\begin{equation}
\bar J = \left(k^\mu(g_{\mu\nu}+b_{\mu\nu})-\omega_\nu\right)\bar\p\Phi^\nu
\end{equation}
vanishes. The discussion proceeds analogously: assuming that $b$ is invariant, $\omega=d\alpha$ is exact, and in a coordinate system $\Phi^0=\alpha$ and $\Phi^i$ such that $\L_k\Phi^i=0$, we have $e_{00}=1$ and $e_{0i}=0$. The formulas (13), (15) and (16) of \cite{Ivanov:1994ec} then show that the metric, torsion and right complex structure are unchanged by T-duality (after a change of coordinates $\tilde\Phi^0\mapsto-\tilde\Phi^0$ of the dual model).

However, T-duality along a left (right) isometry does change the right (left) complex structure. In particular, the structures $J_+\pm J_-$ and $[J_+,J_-]$ are changed. Therefore, T-duality along a Kac-Moody isometry changes the type of the generalized geometry and \emph{relates the different generalized K\"ahler structures} on the same Lie group.

In the following sections, we will perform these T-dualities in $(2,2)$ superspace. Since in $(2,2)$ superspace, gauging an isometry complexifies the gauge group (with respect to both complex structures), T-dualizing along isometries related by the complex structures gives rise to the same T-dual model in superspace. This also implies that isometries related by complex structures cannot be simultaneously gauged \cite{Hull:1985pq}. For the rank 2 Lie groups which we consider in this paper, this means that T-duality along {\em any} Kac-Moody isometry always leads to the same T-dual model.

\subsection{General strategy for finding the generalized K\"ahler potential}
\label{sec.genstrat}
Given the bihermitian data $(g,H,J_+,J_-)$, the generalized K\"ahler potential $K$ can in principle be found by solving the equations \eqref{eqn.gerbefp}-\eqref{eqn.gerbefm}, which relate nonlinearly the Hessian of $K$ to $g$ and $H$ in adapted coordinates. This is a nonlinear second order differential equation -- a difficult equation to solve. However, on the symplectic leaves of $\sigma$ \eqref{eqn.sigma}, there is a simplification. On each symplectic leaf, $K$ generates the symplectomorphism between left holomorphic Darboux coordinates (left semi-chiral superfields) and right holomorphic coordinates (right semi-chiral superfields), which means that it satisfies the \emph{first order linear} differential equations \eqref{eqn.yl1}-\eqref{eqn.yr}.

This simplification, coupled with the observation about T-duality noted above, allows one to find the generalized K\"ahler potentials for all the generalized K\"ahler structures supported by a Lie group admitting a type $(0,0)$ structure. The strategy is as follows. First, we find left and right holomorphic coordinates on the Lie group. This can be done by expanding the left and right invariant frames about the origin, taking the leading term to be given by the holomorphic Lie algebra generators, and solving for the higher order terms order by order using the Maurer-Cartan equations. (See Appendix \ref{app.holcoord} for more details.) Next, for the type $(0,0)$ structure, identify combinations $\ell,\tell$ and $\rx,\trx$ of left and right holomorphic coordinates that are Darboux for $\sigma$. This yields one-form symplectic potentials $\theta_L=\tell\,d\ell+\bar\tell\,d\bar\ell$ and $\theta_R=\rx\,d\trx+\bar\rx\,d\bar\trx$ for the (local) symplectic form $\sigma^{-1}$. The difference $\theta_L-\theta_R$ is closed, and can be integrated to give the generalized K\"ahler potential $K=\int\theta_L-\theta_R$. Next, the potentials for the generalized geometries of other types on the Lie group can be obtained by T-duality, as discussed above.

We illustrate this strategy for $SU(2)\times U(1)$ by first computing the type $(0,0)$ structure, and then T-dualizing along a Kac-Moody isometry to obtain the type $(1,1)$ generalized K\"ahler potential, reproducing a previously known result. We then apply this to $SU(3)$, constructing the type $(1,1)$ generalized K\"ahler potential from the type $(0,0)$ potential.

\section{$SU(2)\times U(1)$}
\setcounter{equation}{0}
The generalized K\"ahler geometry of $SU(2)\times U(1)$ has been studied in detail \cite{Rocek:1991vk,Sevrin:2011mc}. In this section, we revisit these results as a warm up for the $SU(3)$ model.

The outline of the section is as follows. First, we choose complex structures $J_\pm$ on the Lie group and find complex coordinates. As discussed in section \ref{sec.22wzw}, generalized K\"ahler structures of two types are admissible depending on the choice of complex structures. Next, we find generalized K\"ahler potentials for each of the two types. These potentials can be written down in various different forms, differing from one another by generalized K\"ahler transformations and coordinate transformations. Different choices of potentials suit different purposes and the relations between them are illuminating. Finally, we relate the generalized geometries of the two different types by T-duality.

\subsection{Coordinates and generalized K\"ahler potential}
On the Lie algebra, take the basis $\{h,\bar h,e,\bar e\}$ where
\begin{equation}
h=\begin{pmatrix}\zeta&0\\0&-\bar\zeta\end{pmatrix},~~~~~ \bar h=\begin{pmatrix}\bar\zeta&0\\0&-\zeta\end{pmatrix},~~~~~
e=\begin{pmatrix}0&1\\0&0\end{pmatrix},~~~~~
\bar e=\begin{pmatrix}0&0\\1&0\end{pmatrix},
\label{eqn.u2basis}
\end{equation}
$\zeta=\tfrac{1}{2}(1+i),\ \bar\zeta=\tfrac{1}{2}(1-i)$. The two complex structures on the Lie algebra compatible with the choice of Cartan subalgebra $h,\bar h$ are
\begin{equation}
\bJ_1=\operatorname{diag}(i,-i,i,-i),~~~~~ \bJ_2=\operatorname{diag}(-i,i,i,-i).
\end{equation} 
Depending on whether one takes $J_\pm$ induced from the same or from different Lie algebra complex structures $\bJ_1,\bJ_2$, one gets generalized K\"ahler structures of different types on $SU(2)\times U(1)$.

\subsubsection{Type $(0,0)$}
\label{sec.gkp00}
If one takes $J_+$ and $J_-$ both induced from the same Lie algebra complex structure, say $\bJ_1$, then generically the resulting generalized K\"ahler structure has type $(0,0)$; in other words, $[J_+,J_-]$ has full rank at generic points of the group.\footnote{There are loci of positive codimension on which $\ker[J_+,J_-]$ is nontrivial.} In this case, $J_+$ and $J_-$ induce the same orientation on $SU(2)\times U(1)$. The sigma model description is in terms of one set of semi-chiral superfields.

In terms of the group element in the defining representation
\begin{equation}
g = \begin{pmatrix} g_{11} & g_{12} \\ g_{21} & g_{22} \end{pmatrix},
\end{equation}
the $J_\pm$ holomorphic coordinates can be chosen to be\footnote{See Appendix \ref{app.holcoord} for a discussion of how to find such coordinates.}
\begin{equation}
\begin{array}{l} z_+^1 = \log g_{12}^\zeta\bar g_{21}^{\bar\zeta}, \\ z_+^2 = \log g_{22}^\zeta\bar g_{11}^{\bar\zeta}, \\ z_-^1 = \log g_{12}^{\bar\zeta}\bar g_{21}^\zeta, \\ z_-^2 = \log g_{11}^{\bar\zeta}\bar g_{22}^\zeta. \end{array}
\label{eqn.su200param}
\end{equation}
Note that neither the set $(z_+^1,\bar z_+^1,z_-^1,\bar z_-^1)$ nor the set $(z_+^2,\bar z_+^2,z_-^2,\bar z_-^2)$ is nondegenerate (in each case, unitarity of $g$ implies one real relation between these functions). These holomorphic coordinates are chosen to be Darboux with respect to the Poisson structure $\sigma$:
\begin{equation} \sigma(dz_\pm^1,dz_\pm^2)=\pm 1,~~~~~~ \sigma(d\bar z_\pm^1,d\bar z_\pm^2)=\pm 1. \end{equation}
One choice of semi-chiral coordinates is
\begin{equation} \ell=z_+^2,~~~~~ \tell =z_+^1,~~~~~ \rx=z_-^2-z_-^1,~~~~~ \trx =z_-^2, \label{eqn.darb1} \end{equation}
satisfying $d(\tell\ d\ell+\bar\tell\ d\bar\ell+\trx\ d\rx+\bar\trx\ d\bar\rx)=0$. Choosing polarizations such that the adapted coordinates are $\tell,\bar\tell$ and $\rx,\bar\rx $ results in the parametrization \cite{Sevrin:2011mc}
\begin{equation} g = e^{-\zeta\theta}\begin{pmatrix}e^{\tell+\rx} & e^{\tell} \\ -e^{\bar\tell} & e^{\bar\tell+\bar\rx} \end{pmatrix}, ~~~~~\theta=\tell+\bar\tell+\log(1+e^{\rx+\bar \rx }) \label{eqn.klr1param} \end{equation}
and potential
\begin{align}
K^{(0,0)}_0 =& \int -\ell \ d\tell-\bar \ell \ d\bar \tell+\trx\ d\rx +\bar\trx\ d\bar\rx  \nn\\
=& -(\tell+\rx)(\bar\tell+\bar\rx)+\int^{-\rx-\bar\rx }\log(1+e^q)\ dq
\label{eqn.klr1}
\end{align}
satisfying $\frac{\p K}{\p \tell}=-\ell ,\frac{\p K}{\p\rx}=\trx$. This potential is valid on the coordinate patch away from the off-diagonal matrices. On the other coordinate patch, away from the diagonal matrices, we choose the polarizations spanned by $\ell,\bar\ell$ and $\rx,\bar\rx$, which results in the parametrization
\begin{equation}
g = e^{-\zeta\theta}\begin{pmatrix} e^{\bar\ell} & e^{\bar\ell-\rx} \\ -e^{\ell-\bar\rx} & e^{\ell} \end{pmatrix},~~~~~\theta=\ell+\bar\ell+\log(1+e^{-(\rx+\bar\rx)})
\label{eqn.klr1param1}
\end{equation}
and potential
\begin{align}
K^{(0,0)}_1 =& \int \tell \ d\ell+\bar\tell \ d\bar\ell+\trx\ d\rx +\bar\trx\ d\bar\rx  \nn\\
=& (\ell-\bar\rx)(\bar\ell-\rx)-\int^{\rx+\bar\rx}\log(1+e^q)\ dq+\frac{1}{2}(\rx^2+\bar\rx^2)
\label{eqn.klr12}
\end{align}
satisfying $\frac{\p K}{\p\ell}=\tell,\frac{\p K}{\p r}=\trx$. On the overlap of the two patches (comprising the group elements with nonvanishing entries), \eqref{eqn.klr1} and \eqref{eqn.klr12} differ by a Legendre transform
\begin{equation}
K^{(0,0)}_0(\tell,\bar\tell,\rx,\bar\rx) = K_1^{(0,0)}(\ell,\bar\ell,\rx,\bar\rx) - \ell\tell - \bar\ell\bar\tell.
\end{equation}
This appears to be the choice of parametrization and polarization giving the simplest expression for the potential. Some other choices are given in Appendix \ref{sec.appgkp00}.

\subsubsection{Type $(1,1)$}
If one instead takes $J_+$ and $J_-$ induced from different Lie algebra complex structures, say $J_+$ from $\bJ_1$ and $J_-$ from $\bJ_2$, one finds $[J_+,J_-]=0$ everywhere. In other words, the resulting generalized K\"ahler structure has type $(1,1)$, and can be parametrized by a chiral coordinate $\phi$ and a twisted chiral coordinate $\chi$. Furthermore, $J_+$ and $J_-$ induce opposite orientations. In this case, the conditions that $\phi$ is holomorphic with respect to both $J_\pm$ determine it uniquely up to a simple redefinition $\phi\to \phi'(\phi)$. Similarly, $\chi$ is also essentially unique.

One choice of parametrization of the group element is given by
\begin{equation}
g = e^{-\zeta\theta}\begin{pmatrix}e^{\bar\chi}& e^\phi \\ -e^{\bar\phi} & e^\chi\end{pmatrix}, \qquad\qquad \theta=\log(e^{\phi+\bar\phi}+e^{\chi+\bar\chi}).
\label{eqn.bilpparam}
\end{equation}
where we recall that $\zeta=\tfrac{1}{2}(1+i)$. Equivalently, the chiral and twisted chiral coordinates are given by
\begin{equation}
\begin{array}{l} \phi = \log g_{12}^\zeta\bar g_{21}^{\bar\zeta}, \\
\chi = \log g_{22}^\zeta\bar g_{11}^{\bar\zeta}.
\end{array}
\end{equation}
This coordinate patch covers the region where all the entries of the group are nonzero. A redefinition $\hat\phi=e^\phi, \hat\chi=e^\chi$ allows one to reach the diagonal elements ($\hat\phi=0$) and the off-diagonal elements ($\hat\chi=0$).

The generalized potential for the type $(1,1)$ structure is known \cite{Rocek:1991vk}
\begin{equation}
K^{(1,1)}_0 = \frac{1}{2}(\chi-\bar\chi)^2+\int^{{\phi+\bar\phi-\chi-\bar\chi}}dq\ \log(1+e^q).
\label{eqn.kbilp1}
\end{equation}
By redefining $\hat\phi=e^\phi$ and checking that the limit $\hat\phi\to 0$ is well-defined, one can verify that this potential is valid on the coordinate patch away from the off-diagonal matrices.\footnote{In \cite{Rocek:1991vk}, the variables used correspond to $\hat\phi,\hat\chi$.} The following potential
\begin{equation}
K^{(1,1)}_1 = -\frac{1}{2}(\phi-\bar\phi)^2-\int^{{-\phi-\bar\phi+\chi+\bar\chi}}dq\ \log(1+e^q).
\label{eqn.kbilp2}
\end{equation}
is valid on the coordinate patch away from diagonal matrices, as can be seen by redefining $\hat\chi=e^\chi$ and checking that the limit $\hat\chi\to 0$ is well-defined. These two patches cover $SU(2)\times U(1)$. On the overlap of the two patches, comprising the group elements with nonvanishing entries, \eqref{eqn.kbilp1} and \eqref{eqn.kbilp2} differ by a generalized K\"ahler transformation $K^{(1,1)}_0-K^{(1,1)}_1=-(\chi+\bar\chi)(\phi+\bar\phi)$.

The generalized potentials \eqref{eqn.kbilp1}, \eqref{eqn.kbilp2} were obtained by solving the second order differential equations \eqref{eqn.gerbefp}-\eqref{eqn.gerbefm}\footnote{For special case of $SU(2)\times U(1)$, which has no semi-chiral coordinates, these equations turn out to be linear. For generic Lie groups, these equations are nonlinear and difficult to solve.} In the next subsection, we shall make use of the discussion in the previous section to derive the type $(1,1)$ potentials \eqref{eqn.kbilp1}, \eqref{eqn.kbilp2} from the type $(0,0)$ potentials \eqref{eqn.klr1}, \eqref{eqn.klr12} via T-duality.

\subsection{Isometries}
The $SU(2)\times U(1)$ WZW model has isometry group $SU(2)_L\times SU(2)_R\times U(1)$, and the subgroup preserving both complex structures $J_\pm$ is $U(1)_L\times U(1)_R\times U(1)$. It acts on the group element $g$ as
\begin{equation}
g\mapsto h_L g h_R^{-1},
\end{equation}
with
\begin{equation}
h_L = e^{-i(\epsilon h+\bar\epsilon\bar h)},\qquad h_R = e^{i(\eta\bar h+\bar\eta h)},
\end{equation}
where $h$ is defined in \eqref{eqn.u2basis}, and $\epsilon$ and $\eta$ are complex parameters.

In the type $(0,0)$ parametrization \eqref{eqn.darb1}, this corresponds to
\begin{equation}
\begin{array}{ll}
\ell \mapsto \ell+\bar\epsilon+\eta,~~~~~&\tell\mapsto\tell+\epsilon+\eta \\
\rx \mapsto \rx-\eta+\bar\eta, &\trx\mapsto\trx-\bar\epsilon-\eta.
\end{array}
\label{eqn.isom00}
\end{equation}
The parameters $\epsilon$ and $\eta$ can be promoted to Kac-Moody parameters satisfying
\begin{equation}
\begin{array}{l}
\bar {\bD}_+\epsilon = 0,\ {\bD}_\pm\epsilon=0, \\
\bar {\bD}_\pm\eta = 0,\ {\bD}_-\eta=0.
\end{array}
\end{equation}

In the type $(1,1)$ parametrization \eqref{eqn.bilpparam}, this corresponds to
\begin{equation}
\phi \mapsto \phi+\epsilon+\eta,~~~~~ \chi\mapsto\chi+\bar\epsilon+\eta.
\label{eqn.isom11}
\end{equation}
The parameters $\epsilon$ and $\eta$ can be promoted to Kac-Moody parameters satisfying (note the chirality constraints on $\epsilon$ differ from above)
\begin{equation}
\begin{array}{l}
\bar {\bD}_\pm\epsilon = 0,\ {\bD}_+\epsilon=0, \\
\bar {\bD}_\pm\eta=0,\ {\bD}_-\eta=0.
\end{array}
\end{equation}

\subsection{T-duality: Relating the two types}
As discussed in section \ref{sec.wzwtdual}, T-duality along a Kac-Moody isometry relates the two generalized structures on $SU(2)\times U(1)$. This T-duality may be realized in $(2,2)$ superspace using the gauging prescription of \cite{Rocek:1991ps}.

\subsubsection{Type $(0,0)$ to type $(1,1)$}
We begin with the type $(0,0)$ generalized K\"ahler structure \eqref{eqn.klr1}, and T-dualize along any factor of the $U(1)_L\times U(1)_R\times U(1)$ Kac-Moody isometry group. The complex structures $J_\pm$ map these isometries into one another, so in superspace, where the gauge group is complexified, the gauging of any of these isometries is equivalent (up to reparametrizations).

\vspace{1ex}
\noindent{\underline{Along $U(1)_R$}}

Consider first T-duality along $U(1)_R$, which acts on the semi-chiral coordinates as in \eqref{eqn.isom00} with $\epsilon=0,\eta=i\lambda$, where $\lambda$ is a real parameter. This isometry is gauged with an Semichiral Vector Multiplet (SVM) \cite{Lindstrom:2007vc} (see Appendix \ref{sec.tduality}). The combinations invariant under the $U(1)_R$ isometry are $\tell+\bar\tell$, $-\tfrac{1}{2}(\rx+\bar\rx)$ and $i(\bar\tell-\tell+\tfrac{1}{2}(\bar\rx-\rx))$, and are respectively gauged with the potentials $V^L$, $V^R$ and $V'$ of the SVM. Starting from $K^{(0,0)}_0$ in \eqref{eqn.klr1}, we add the generalized K\"ahler transformation term
\begin{equation}
K^{(0,0)} = K^{(0,0)}_0 +\frac{1}{2}(\tell^2+\bar\tell^2)-\frac{1}{4}(\rx^2+\bar\rx^2)
\label{eqn.k00gkterms}
\end{equation}
to make the potential exactly invariant.\footnote{This is not necessary but simplifies the discussion somewhat.} The generalized K\"ahler transformation terms amount to redefining $\ell$ and $\trx$ to the invariant combinations $\ell\mapsto\ell-\tell$, $\trx\mapsto\trx-\rx/2$. The resulting invariant potential is
\begin{equation}
K^{(0,0)} = \frac{1}{2}(\bar\tell-\tell+\tfrac{1}{2}(\bar\rx-\rx))^2-\frac{3}{8}(\rx+\bar\rx)^2-\frac{1}{2}(\tell+\bar\tell)(\rx+\bar\rx)+\int^{-\rx-\bar\rx}dq\ \log(1+e^q).
\end{equation}
Gauging with an SVM, enforced to be flat by Lagrange multipliers $\Phi_I$, and gauge fixing $\tell=\bar\tell=\rx=\bar\rx=0$ yields
\begin{equation}
\tilde K^{(0,0)} = -\frac{1}{2}(V')^2 - \frac{3}{2}(V^R)^2 + V^LV^R+\int^{2V^R}dq\ \log(1+e^q) - V^I\Phi_I,
\end{equation}
where the Lagrange multipliers are
\begin{equation}\begin{array}{rl}
\Phi_L &=\tfrac{1}{2}(\phi+\bar\phi-\chi-\bar\chi), \\
\Phi_R &=\tfrac{1}{2}(-\phi-\bar\phi-\chi-\bar\chi), \\
\Phi' &=\tfrac{i}{2}(\phi-\bar\phi+\bar\chi-\chi).
\end{array}\end{equation}
Eliminating the SVM gauge fields yields the T-dual potential
\begin{equation}
\tilde K^{(0,0)} = -\frac{1}{2}(\chi+\bar\chi)^2 + \int^{\phi+\bar\phi-\chi-\bar\chi}dq\ \log(1+e^q) - \frac{1}{2}\left((\phi-\chi)^2+(\bar\phi-\bar\chi)^2\right)+\frac{1}{4}\left((\phi+\bar\chi)^2+(\bar\phi+\chi)^2\right),\end{equation}
which is precisely $K^{(1,1)}_0$ \eqref{eqn.kbilp1} up to a generalized K\"ahler transformation.

\vspace{1ex}
\noindent{\underline{Along $U(1)_L$}}

Consider now T-duality along the $U(1)_L$ factor of $U(1)_L\times U(1)_R\times U(1)$, which acts on the coordinates as in \eqref{eqn.isom00} with $\epsilon=i\lambda,\eta=0$, where $\lambda$ is a real parameter. In this case, there is a complication because $\rx,\bar\rx$ are invariant - how does the potential \eqref{eqn.klr1} couple to the SVM in this case? One way to do so is to perform a Legendre transform from $\rx,\bar\rx$ to $\trx,\bar\trx$ (corresponding to a change in polarization), yielding the potential (we also add the generalized K\"ahler transformation $-\tfrac{1}{2}(\tell^2+\bar\tell^2)$ to render the potential invariant - this amounts to redefining $\ell\to\ell+\tell$)
\begin{align}
K^{(0,0)}(\tell,\bar\tell,\trx,\bar\trx) &= K^{(0,0)}_0(\tell,\bar\tell,\rx,\bar\rx) -\frac{1}{2}(\tell^2+\bar\tell^2)- \rx\trx - \bar\rx\bar\trx \nn \\
&= -\frac{1}{2}(\tell+\bar\tell)^2 - \rx(\bar\tell+\trx)-\bar\rx(\tell+\bar\trx)-\rx\bar\rx + \int^{-\rx-\bar\rx}dq\ \log(1+e^q).
\label{eqn.klrinv}
\end{align}
Note that $\trx$ does indeed transform as $\trx\mapsto\trx+i\lambda$, so now the invariant combinations $\tell+\bar\tell$, $\tell+\bar\trx$ and $\bar\tell+\trx$ can be respectively gauged with the components $V^L$, $\tilde\bV$ and $\bar{\tilde\bV}$ of the SVM. The T-dual potential is
\begin{equation}
\tilde K^{(0,0)} = -\frac{1}{2}(V^L)^2 -\rx\bar\rx - \rx\bar{\tilde\bV} - \bar\rx\tilde\bV + \int^{-\rx-\bar\rx}dq\ \log(1+e^q) - V^I\Phi_I,
\end{equation}
where
\begin{equation}
V^I\Phi_I = V^Li(\bar\phi-\phi)+\tilde\bV(\bar\phi-\chi)+\bar{\tilde\bV}(\phi-\bar\chi).
\end{equation}
Note that the variational equations of $V^I$ are
\begin{equation}
0=\left(\frac{\p K^{(0,0)}_0}{\p\rx}+\trx\right)\frac{\p\rx}{\p V^I} + \left(\frac{\p K^{(0,0)}_0}{\p\bar\rx}+\bar\trx\right)\frac{\p\bar\rx}{\p V^I} + \frac{\p K^{(0,0)}}{\p V^I},
\end{equation}
where the derivative in the last term is taken with $\rx,\bar\rx$ held fixed. The two terms in parentheses vanish. In particular, the variational equation of $\tilde\bV$ sets $\bar\rx=\chi-\bar\phi$. The T-dual potential then simplifies to
\begin{equation}
\tilde K^{(0,0)} = \frac{1}{2}(\chi-\bar\chi)^2 + \int^{\phi+\bar\phi-\chi-\bar\chi}dq\ \log(1+e^q) - \frac{1}{2}\left((\phi-\chi)^2+(\bar\phi-\bar\chi)^2\right),
\label{eqn.klrinvtilde}
\end{equation}
which we recognize as \eqref{eqn.kbilp1} up to a generalized K\"ahler transformation. We have obtained the type $(1,1)$ potential without the need to solve second order differential equations.

\subsubsection{Using the group coordinates to find the T-dual}
\label{sec.u2obs}
We make an observation of the T-dualities which we performed, which we will apply to simplify the discussion in the $SU(3)$ case.

In the type $(0,0)$ structure, both complex structures $J_\pm$ were induced from the Lie algebra structure $\bJ_1$. As discussed in section \ref{sec.wzwtdual}, T-duality along the left Kac-Moody isometry $U(1)_L$ does not change the metric, torsion and left complex structure. Therefore, on the dual type $(1,1)$ structure, we know exactly what the complex structures are: $J_+$ is unchanged and is still induced from $\bJ_1$, while $J_-$ is changed and is now induced from $\bJ_2$. The adapted coordinates for this particular generalized K\"ahler structure are already known and given in \eqref{eqn.bilpparam}. The solution to the SVM equations of motion must therefore be given by the original type $(0,0)$ coordinates \eqref{eqn.darb1},\footnote{with $\ell$ replaced with $\ell+\tell$ corresponding to the addition of the generalized K\"ahler transformation term $-\tfrac{1}{2}(\tell^2+\bar\tell^2)$} expressed in terms of the type $(1,1)$ coordinates of the dual \eqref{eqn.bilpparam}, to wit
\begin{equation}
\begin{array}{ll} \ell=\chi+\phi,\qquad &\tell =\phi, \\ \rx=\bar\chi-\phi,\qquad &\trx =\phi-\theta, \end{array}\qquad\qquad \text{ where }\theta=\log(e^{\phi+\bar\phi}+e^{\chi+\bar\chi}).
\label{eqn.coord1}
\end{equation}

Notice also that, when the isometries act as translations (all the examples we encounter in this paper are translational isometries), the Lagrange multiplier term may be written as
\begin{equation}
V^I\Phi_I = -\ell(\Phi)\, \tell-\bar\ell(\Phi)\,\bar\tell-\rx(\Phi)\,\bar\rx-\bar\rx(\Phi)\,\bar\trx,
\end{equation}
where $\ell(\Phi),\bar\ell(\Phi),\rx(\Phi),\bar\rx(\Phi)$ are functions of $\Phi=(\phi,\bar\phi,\chi,\bar\chi)$ given in \eqref{eqn.coord1}. This yields a simple integral expression for the type $(1,1)$ potential
\begin{align}
K^{(1,1)}(\phi,\bar\phi,\chi,\bar\chi) &= \left(-\int \ell\, d\tell + \bar\ell\, d\bar\tell + \rx\, d\trx + \bar\rx\, d\bar\trx\right) - V^I\Phi_I \nn \\
&= \int \tell\, d\ell + \bar\tell\, d\bar\ell + \trx\, d\rx + \bar\trx\, d\bar\rx,
\end{align}
where the semi-chiral fields $\ell,\tell,\rx,\bar\rx$ are to be understood as functions of $\phi,\chi$ through \eqref{eqn.coord1}. It is straightforward to verify, via direct substitution, that this exactly reproduces \eqref{eqn.klrinvtilde}.

For another illustration of using the group coordinates to solve the vector multiplet moment map equations, see Appendix \ref{sec.su2u1tdual} where the T-duality in the other direction, which is done with an Large Vector Multiplet (LVM) \cite{Lindstrom:2007vc}, is discussed.
 
\section{$SU(3)$}
\setcounter{equation}{0}
\subsection{Complex coordinates and generalized K\"ahler potential}
On the Lie algebra, take the basis $\{h,\bar h,e_3,\bar e_3,e_1,\bar e_1,e_2,\bar e_2\}$, where
\begin{equation}
h=\begin{pmatrix}\frac{1}{2}+\frac{i}{2\sqrt{3}}&&\\&-\frac{i}{\sqrt{3}}&\\&&-\frac{1}{2}+\frac{i}{2\sqrt{3}}\end{pmatrix},\ e_3=\begin{pmatrix}\ &\ &1\\&&\\&&\end{pmatrix},\ e_1=\begin{pmatrix}\ &1&\ \\&&\\&&\end{pmatrix},\ e_2=\begin{pmatrix}\ &\ &\ \\&&1\\&&\end{pmatrix},
\label{eqn.algbasis}
\end{equation}
and the bar denotes hermitian conjugation. The two complex structures on the Lie algebra compatible with the choice of Cartan subalgebra $h,\bar h$ are
\begin{equation}
\mathbb{J}_1=\operatorname{diag}(i,-i,i,-i,i,-i,i,-i),\qquad \text{ and }\qquad\mathbb{J}_2=\operatorname{diag}(-i,i,i,-i,i,-i,i,-i).
\end{equation}
As discussed in section \ref{sec.22wzw}, $SU(3)$ admits generalized K\"ahler structures of two different types depending on whether one takes $J_\pm$ induced from the same or from different Lie algebra complex structures $\bJ_1,\bJ_2$.

\subsubsection{Type $(0,0)$}
If one takes $J_+$ and $J_-$ both to be induced from the same Lie algebra complex structure, say $\mathbb{J}_1$, then generically the resulting generalized K\"ahler structure has type $(0,0)$.\footnote{There are loci of positive codimension on which $\ker[J_+,J_-]$ is nontrivial.} The $J_\pm$ complex coordinates take the simplest form when presented in an overcomplete basis,
\begin{align}
z_+^\phi =& \log \bar g_{31}^{\bar\omega}g_{13}^\omega,\qquad z_+^\chi = \log\bar g_{11}^{\bar\omega}g_{33}^\omega,\nonumber \\
z_+^1 =& \log\frac{g_{13}}{g_{23}}, \qquad z_+^2 = \log\frac{g_{23}}{g_{33}},\qquad z_+^3 = \log\frac{\bar g_{11}}{\bar g_{21}}, \qquad z_+^4 = \log\frac{\bar g_{21}}{\bar g_{31}}, \\
z_-^\phi =& \log\bar g_{31}^\omega g_{13}^{\bar\omega},\qquad z_-^{\bar\chi} = \log g_{11}^{\bar\omega}\bar g_{33}^\omega,\nonumber \\ z_-^1 =& \log\frac{g_{11}}{g_{12}},\qquad z_-^2 = \log\frac{g_{12}}{g_{13}}, \qquad z_-^3 = \log\frac{\bar g_{31}}{\bar g_{32}},\qquad z_-^4 = \log\frac{\bar g_{32}}{\bar g_{33}},
\end{align}
where $g_{ij}$ is the $(i,j)$th entry of the group element $g$ in the defining representation of $SU(3)$, and $\omega=e^{i\pi/3}=\frac{1}{2}(1+i\sqrt{3})$. For each of $+$ and $-$, the six complex coordinates satisfy two relations
\begin{align}
&e^{z_\pm^1+z_\pm^3}+e^{-z_\pm^2-z_\pm^4}+1=0, \nn\\
&z_+^\phi-z_+^\chi = \omega(z_+^1+z_+^2)-\bar\omega(z_+^3+z_+^4), \\
&z_-^\phi-z_-^{\bar\chi} = -\bar\omega(z_-^1+z_-^2)+\omega(z_-^3+z_-^4). \nn
\end{align}

Semichiral coordinates are holomorphic coordinates that are Darboux with respect to the Poisson structure $\sigma$: $\sigma(d\ell^j,d\tell^k)=\delta^{jk}$, $\sigma(d\ell^j,d\ell^k)=\sigma(d\tell^j,d\tell^k)=0$, $\sigma(d\rx^j,d\trx^k)=-\delta^{jk}$ and $\sigma(d\rx^j,d\rx^k)=\sigma(d\trx^j,d\trx^k)=0$. One choice of semi-chiral coordinates is given by
\begin{equation}
\begin{array}{l@{\,}l@{\,}ll}
\hat\ell^1 &=& \frac{1}{3}(z_+^\chi+2z_+^\phi-\omega z_+^1+\bar\omega z_+^4) &= \frac{1}{3}\log (\bar g_{11}\bar g_{21}\bar g_{31})^{\bar\omega}(g_{13}g_{23}g_{33})^\omega, \\
\hat\tell^1 &=& (\bar\omega-\omega)(z_+^1+z_+^2+z_+^3+z_+^4) &= \displaystyle (\bar\omega-\omega)\log\frac{\bar g_{11}g_{13}}{\bar g_{31}g_{33}}, \\
\hat\ell^2 &=& z_+^\chi &= \log \bar g_{11}^{\bar\omega}g_{33}^\omega, \\
\hat\tell^2 &=& z_+^\phi &= \log \bar g_{31}^{\bar\omega}g_{13}^\omega, \\
\hat\rx^1 &=&  (\bar\omega-\omega)(z_-^1+z_-^2+z_-^3+z_-^4) &= \displaystyle (\bar\omega-\omega)\log\frac{g_{13}\bar g_{33}}{g_{11}\bar g_{31}}, \\
\hat\trx^1 &=& \frac{1}{3}(z_-^{\bar\chi}+2z_-^\phi+\bar\omega z_-^2-\omega z_-^3) &= \frac{1}{3}\log (g_{11}g_{12}g_{13})^{\bar\omega}(\bar g_{31}\bar g_{32}\bar g_{33})^\omega, \\
\hat\rx^2 &=& z_-^{\bar\chi} &= \log g_{11}^{\bar\omega}\bar g_{33}^\omega, \\
\hat\trx^2 &=& z_-^\phi &= \log g_{13}^{\bar\omega}\bar g_{31}^\omega.
\end{array}
\end{equation}
(Recall $\omega=e^{i\pi/3}=\frac{1}{2}(1+i\sqrt{3})$.) Choosing the polarizations defined by $\hat\ell^j,\bar{\hat\ell}^j$ and $\hat\trx^j,\bar{\hat\trx}^j$, the group element is parametrized as
\begin{equation}{\scriptsize
g = \begin{pmatrix}
\exp(\bar{\hat\ell}^2+u\hat h_1) &
\exp(3\hat\trx^1-\bar{\hat\ell}^2-\hat\trx^2-\hat h_1+\bar u\hat h_4)&
\exp(\hat\trx^2-\bar u\hat h_2) \\
-\exp(3\bar{\hat\ell}^1-\bar{\hat\ell}^2-\bar{\hat\trx}^2-\bar {\hat h}_2+u\hat h_3) &
g_{22} &
\exp(3\hat\ell^1-\hat\ell^2-\hat\trx^2-\hat h_2+u\bar{\hat h}_3) \\
\exp(\bar{\hat\trx}^2-\bar u\bar {\hat h}_2) &
-\exp(3\bar{\hat \trx}^1-\bar{\hat\trx}^2-\hat\ell^2-\bar {\hat h}_1+\bar u\bar{\hat h}_4)&
\exp(\hat\ell^2+u\bar {\hat h}_1)
\end{pmatrix}}
\end{equation}
where $u=\tfrac{1}{\sqrt{3}}e^{i\pi/6}=\frac{1}{2}(1+\frac{i}{\sqrt{3}})$ and $\hat h_1,\hat h_2,\hat h_3,\hat h_4$ are functions of $\hat \ell^j,\bar{\hat\ell}^j,\hat\trx^j,\bar{\hat\trx}^j$, determined by the orthonormality of the first and third rows and columns\footnote{Unfortunately, it seems that $\hat h_1,\hat h_2,\hat h_3,\hat h_4$ cannot be written down in terms of elementary functions, so we have to work implicitly.}; and finally $g_{22}$ may be determined by unimodularity.

To find the potential, we need to express $\hat\tell^j,\hat\rx^j$ in terms of $\hat\ell^j,\bar{\hat\ell}^j,\hat\trx^j,\bar{\hat\trx}^j$. It is straightforward to verify that
\begin{equation}
\begin{array}{l}
\hat\tell^1 = -\bar{\hat h}_1+\hat h_2, \\
\hat\tell^2 = \hat\trx^2-\hat h_2, \\
\hat\rx^1 = -\hat h_1+\hat h_2, \\
\hat\rx^2 = \bar{\hat\ell}^2+\hat h_1.
\end{array}
\end{equation}
The difference between the one-form symplectic potentials adapted to the left and right semi-chiral coordinates is the closed one-form $\hat\theta_1+\hat\theta_2$, where
\begin{equation}
\hat\theta_j= \hat\tell^jd\hat\ell^j+\bar{\hat\tell}^jd\bar{\hat\ell}^j-\hat\rx^jd\hat\trx^j-\bar{\hat\rx}^jd\bar{\hat\trx}^j
\qquad\text{ (no sum over $j$).}
\end{equation}
The potential is given by
\begin{equation}
K = \int_{\mathcal{O}}^{(\hat\ell^j,\bar{\hat\ell}^j,\hat\trx^j,\bar{\hat\trx}^j)}\hat\theta_1+\hat\theta_2,
\end{equation}
where $\mathcal{O}$ is some base point. The closure condition $d(\hat\theta_1+\hat\theta_2)=0$ ensures that the integral is independent of path. $K$ generates the symplectomorphism between left and right semi-chiral coordinates
\begin{equation}
\begin{array}{ll}
\ds\frac{\p K}{\p\hat\ell^j} = \hat\tell^j,\qquad&\ds \frac{\p K}{\p\bar{\hat\ell}^j}=\bar{\hat\tell}^j,\\[10pt]
\ds\frac{\p K}{\p\hat\trx^j} = -\hat\rx^j,\qquad&\ds\frac{\p K}{\p\bar{\hat\trx}^j}=-\bar{\hat\rx}^j.
\end{array}
\label{eqn.ylyr}
\end{equation}

\subsubsection*{A different choice}
In section \ref{sec.su3tdual}, it will prove convenient to choose instead the following Darboux coordinates (these are adapted to the isometry considered later in section \ref{sec.su3tdual})
\begin{equation}
\begin{array}{l@{\,}l@{\,}ll}
\ell^1 &=& \frac{1}{3}(z_+^\chi+2z_+^\phi-\omega z_+^1+\bar\omega z_+^4) &= \frac{1}{3}\log (\bar g_{11}\bar g_{21}\bar g_{31})^{\bar\omega}(g_{13}g_{23}g_{33})^\omega, \\[2mm]
\tell^1 &=& (\bar\omega-\omega)(z_+^1+z_+^2+z_+^3+z_+^4) &= \displaystyle (\bar\omega-\omega)\log\frac{\bar g_{11}g_{13}}{\bar g_{31}g_{33}}, \\[1mm]
\ell^2 &=& z_+^\chi &= \log \bar g_{11}^{\bar\omega}g_{33}^\omega, \\[1mm]
\tell^2 &=& z_+^\phi+z_+^\chi &= \log (\bar g_{11}\bar g_{31})^{\bar\omega}(g_{13}g_{33})^\omega, \\[1mm]
\rx^1 &=& (\omega-2\bar\omega)(z_-^1+z_-^2)+(2\omega-\bar\omega)(z_-^3+z_-^4)  &= \displaystyle (2\bar\omega-\omega)\log\frac{g_{13}}{g_{11}}+(\bar\omega-2\omega)\log\frac{\bar g_{33}}{\bar g_{31}}, \\[3mm]
\trx^1 &=& \frac{1}{6}(z_-^{\bar\chi}-z_-^\phi+\bar\omega z_-^2-\omega z_-^3) &= \displaystyle\frac{1}{6}\log \left(\frac{g_{11}g_{12}}{g_{13}^2}\right)^{\bar\omega}\left(\frac{\bar g_{32}\bar g_{33}}{\bar g_{31}^2}\right)^\omega, \\[4mm]
\rx ^2 &=& \omega(z_-^1+z_-^2)-\bar\omega(z_-^3+z_-^4) &= \displaystyle\log\frac{g_{11}^\omega \bar g_{33}^{\bar\omega}}{g_{13}^\omega\bar g_{31}^{\bar\omega}}, \\[2mm]
\trx^2 &=& \frac{1}{6}(z_-^{\bar\chi}+5z_-^\phi+\bar\omega z_-^2-\omega z_-^3) &= \frac{1}{6}\log (g_{11}g_{12}g_{13}^4)^{\bar\omega}(\bar g_{31}^4\bar g_{32}\bar g_{33})^\omega.
\end{array}
\label{eqn.coordsemi}
\end{equation}
In the polarizations defined by $\ell^j,\bar\ell^j$ and $\trx^j,\bar\trx^j$, the group element is parametrized as

\begin{equation}
{\scriptsize
g=\begin{pmatrix}
\exp(\bar\ell^2+uh_1) &
\exp(4\trx^1-\bar\ell^2+2\trx^2-h_1+\bar uh_4)&
\exp(-\trx^1+\trx^2-\bar uh_2) \\
-\exp(3\bar\ell^1-\bar\ell^2+\bar\trx^1-\bar\trx^2-\bar h_2+uh_3) &
g_{22} &
\exp(3\ell^1-\ell^2+\trx^1-\trx^2-h_2+u\bar h_3) \\
\exp(-\bar\trx^1+\bar\trx^2-\bar u\bar h_2) &
-\exp(4\bar\trx^1+2\bar\trx^2-\ell^2-\bar h_1+\bar u\bar h_4)&
\exp(\ell^2+u\bar h_1)
\end{pmatrix},}
\label{eqn.00param}
\end{equation}
\vskip1mm
\noindent where $h_1,h_2,h_3,h_4$ are once again functions of $\ell^j,\bar\ell^j,\trx^j,\bar\trx^j$ determined by the orthonormality of the first and last rows and columns. The Darboux partners are expressed in this mixed coordinate system as
\begin{equation}
\begin{array}{l}
\tell^1 = -\bar h_1+h_2, \\
\tell^2 = -\trx^1+\trx^2+\ell^2-h_2, \\
\rx^1 = -\bar\ell^2-\trx^1+\trx^2-2h_1+h_2, \\
\rx^2 = \bar\ell^2+\trx^1-\trx^2+h_2.
\end{array}
\label{eqn.00y}
\end{equation}
The generalized potential is given by
\begin{align}
&K^{(0,0)} = \int_{\mathcal{O}}^{(\ell^j,\bar\ell^j,\trx^j,\bar\trx^j)}\theta_1+\theta_2,\qquad\text{where }\theta=\theta_1+\theta_2,
\label{eqn.00pot} \\
&\theta_j= \tell^jd\ell^j+\bar\tell^jd\bar\ell^j-\rx^jd\trx^j-\bar\rx^jd\bar\trx^j
\qquad\text{ (no sum over $j$)}
\end{align}

\subsubsection{Type $(1,1)$}
If one instead takes $J_+$ and $J_-$ induced from different Lie algebra complex structures, say $J_+$ from $\mathbb{J}_1$ and $J_-$ from $\mathbb{J}_2$, then generically the resulting generalized K\"ahler structure has type $(1,1)$. The biholomorphic coordinate is
\begin{equation}
\phi = \log\bar g_{31}^{\bar\omega}g_{13}^\omega,
\label{eqn.coordphi}
\end{equation}
and represents a chiral superfield in the $(2,2)$ sigma model, while the $J_+$-holomorphic and $J_-$-antiholomorphic coordinate is
\begin{equation}
\chi = \log\bar g^{\bar\omega}_{11}g_{33}^\omega,
\label{eqn.coordchi}
\end{equation}
and represents a twisted chiral superfield. The other holomorphic coordinates once again take the simplest form in an overcomplete basis
\begin{align}
w_+^1 =& \log\frac{g_{13}}{g_{23}}, \qquad w_+^2 = \log\frac{g_{23}}{g_{33}},\qquad w_+^3 = \log\frac{\bar g_{11}}{\bar g_{21}}, \qquad w_+^4 = \log\frac{\bar g_{21}}{\bar g_{31}}, \\
w_-^1 =& \log\frac{g_{11}}{g_{12}},\qquad w_-^2 = \log\frac{g_{12}}{g_{13}}, \qquad w_-^3 = \log\frac{\bar g_{31}}{\bar g_{32}},\qquad w_-^4 = \log\frac{\bar g_{32}}{\bar g_{33}},
\end{align}
satisfying
\begin{align}
&e^{w_\pm^1+w_\pm^3}+e^{-w_\pm^2-w_\pm^4}+1=0, \\
&\phi-\chi=\omega(w_+^1+w_+^2)-\bar\omega(w_+^3+w_+^4),\\
&\phi-\bar\chi=-\omega(w_-^1+w_-^2)+\bar\omega(w_-^3+w_-^4).
\end{align}
One choice of semi-chiral coordinates is
\begin{equation}
\begin{array}{l@{\,}l@{\,}ll}
\hat\ell &=& \phi-\omega w_+^1+\bar\omega w_+^4 &= \log g_{23}^\omega\bar g_{21}^{\bar\omega}, \\
\hat\tell  &=& -w_+^1-w_+^2+w_+^3+w_+^4 &= \displaystyle\log \frac{\bar g_{11}g_{33}}{\bar g_{31}g_{13}}, \\
\hat\rx  &=& -w_-^1-w_-^2+w_-^3+w_-^4 &= \displaystyle\log \frac{\bar g_{31}g_{13}}{g_{11}\bar g_{33}}, \\
\hat\trx &=& \phi+\omega w_-^2-\bar\omega w_-^3 &= \log g_{12}^\omega\bar g_{32}^{\bar\omega}.
\end{array}
\end{equation}
In the polarizations defined by $\hat\ell,\bar{\hat\ell}$ and $\hat\trx,\bar{\hat\trx}$, the parametrization looks relatively uncluttered
\begin{equation}
g = \begin{pmatrix} e^{\bar\chi+u\bar{\hat f}_1} & e^{\hat\trx+u\hat f_4} & e^{\phi+u\hat f_2} \\
-e^{\bar{\hat\ell}+u\bar{\hat f}_3} & g_{22} & e^{\hat\ell+u\hat f_3} \\
e^{\bar\phi+u\bar{\hat f}_2} & -e^{\bar{\hat\trx}+u\bar{\hat f}_4} & e^{\chi+u\hat f_1} \end{pmatrix}.
\end{equation}
(Recall $u=\tfrac{1}{\sqrt{3}}e^{i\pi/6}=\frac{1}{2}(1+\frac{i}{\sqrt{3}})$.) The conditions of orthonormality on the first and third rows and columns are
\begin{equation}
\begin{array}{ll}
e^{\phi+\chi}(e^{\bar u\hat f_1+u\hat f_2}+e^{u\hat f_1+\bar u\hat f_2})-e^{2\hat\ell+\sqrt{3}\hat f_3} &= 0, \\
e^{\phi+\bar\chi}(e^{\bar u\bar{\hat f}_1+u\hat f_2}+e^{u\bar{\hat f}_1+\bar u\hat f_2})-e^{2\hat\trx+\sqrt{3}\hat f_4} &= 0, \\
e^{2\re(\chi+u\hat f_1)}+e^{2\re(\phi+u\hat f_2)}+e^{2\re(\hat\ell+u\hat f_3)} &= 1, \\
e^{2\re(\chi+\bar u\hat f_1)}+e^{2\re(\phi+\bar u\hat f_2)}+e^{2\re(\hat\ell+\bar u\hat f_3)}&=1, \\
e^{2\re(\chi+\bar u\hat f_1)}+e^{2\re(\phi+u\hat f_2)}+e^{2\re(\hat\trx+u\hat f_4)}&=1, \\
e^{2\re(\chi+u\hat f_1)}+e^{2\re(\phi+\bar u\hat f_2)}+e^{2\re(\hat\trx+\bar u\hat f_4)}&=1.
\end{array}
\label{eqn.oneqns}
\end{equation}
The first equation of \eqref{eqn.oneqns} is complex, and may be solved for $\hat f_3$ in terms of $\phi,\chi,\hat\ell,\hat f_1,\hat f_2$. This may then be substituted into the third and fourth equations, which are real. Similarly, the second complex equation may be solved for $\hat f_4$ in terms of $\phi,\bar\chi,\hat\trx,\bar{\hat f}_1,\hat f_2$, and substituted into the fifth and sixth equations. This yields four real equations, from which we may solve for $\hat f_1,\hat f_2,\bar{\hat f}_1$ and $\bar{\hat f}_2$.\footnote{As before, it is not possible to write down $\hat f_1,\hat f_2,\bar{\hat f}_1,\bar{\hat f}_2$ using elementary functions, so we work implicitly.} It is not obvious, at first glance, that these 8 real equations \eqref{eqn.oneqns} are independent -- for instance, the equations $gg^\dag=\mathbbm{1}$ and $g^\dag g=\mathbbm{1}$ are equivalent -- but we have checked that they indeed are, and hence uniquely determine the $\hat f_i,\bar{\hat f}_i$s.

Note that these equations \eqref{eqn.oneqns} exhibit two involutive symmetries. There is first a left-right symmetry given by exchanging $\chi\leftrightarrow\bar\chi,\hat\ell\leftrightarrow\hat\trx,\hat f_1\leftrightarrow\bar{\hat f}_1,\hat f_3\leftrightarrow\hat f_4$, which exchanges the first, third, fourth equations with the second, fifth and sixth equations. Under this symmetry, the Darboux partners
\begin{align}
\hat\tell =& \log\frac{\bar g_{11}g_{33}}{\bar g_{31}g_{13}} = 2\chi-2\phi+\hat f_1-\hat f_2, 
\label{eqn.yl} \\
\hat\rx =& \log\frac{\bar g_{31}g_{13}}{g_{11}\bar g_{33}} = 2\phi-2\bar\chi-\bar{\hat f}_1+\hat f_2
\label{eqn.y}
\end{align}
are interchanged with a twist, $\hat\tell\leftrightarrow-\hat\rx $. Second, there is a local mirror symmetry which exchanges $\phi\leftrightarrow\chi,\hat\trx\leftrightarrow\bar{\hat\trx},\hat f_1\leftrightarrow \hat f_2,\hat f_4\leftrightarrow\bar{\hat f}_4,\hat\tell\leftrightarrow-\hat\tell,\hat\rx\leftrightarrow-\bar{\hat\rx}$. This exchanges the second equation with its conjugate, and the fifth with the sixth equation, and preserves the other equations.

On top of symplectomorphisms on each symplectic leaf, the semi-chiral coordinates may also be redefined by arbitrary functions of chiral and twisted chiral coordinates (with the appropriate holomorphy)
\begin{equation}\begin{array}{l}
\hat\ell\mapsto\hat\ell'(\hat\ell,\hat\tell,\phi,\chi), \\
\hat\trx\mapsto\hat\trx'(\hat\rx,\hat\trx,\phi,\bar\chi).
\end{array}\end{equation}
There is therefore a large amount of freedom in the choice of holomorphic Darboux coordinates $\hat\ell,\hat\tell,\hat\rx,\hat\trx$.

\subsubsection*{A different choice}
In section \ref{sec.su3tdual} we will find it convenient to use a different choice of semi-chiral coordinates (adapted to an isometry introduced later in section \ref{sec.su3tdual}), given by
\begin{equation}
\begin{array}{l@{\,}l@{\,}ll}
\ell &=& \frac{1}{3}(w_+^\chi+2w_+^\phi-\omega w_+^1+\bar\omega w_+^4) &= \frac{1}{3}\log (\bar g_{11}\bar g_{21}\bar g_{31})^{\bar\omega}(g_{13}g_{23}g_{33})^\omega, \\
\tell  &=& (\bar\omega-\omega)(w_+^1+w_+^2+w_+^3+w_+^4) &= \displaystyle (\bar\omega-\omega)\log\frac{\bar g_{11}g_{13}}{\bar g_{31}g_{33}}, \\
\rx  &=&  (\omega-2\bar\omega)(w_-^1+w_-^2) + (2\omega-\bar\omega)(w_-^3+w_-^4) &= \displaystyle (\omega-2\bar\omega)\log\frac{g_{11}}{g_{13}}+(\bar\omega-2\omega)\log\frac{\bar g_{33}}{\bar g_{31}}, \\
\trx &=& \frac{1}{6}(\bar\omega w_-^1+2\bar\omega w_-^2-2\omega w_-^3-\omega w_-^4) &= \displaystyle\frac{1}{6}\log \left(\frac{g_{11}g_{12}}{g_{13}^2}\right)^{\bar\omega}\left(\frac{\bar g_{32}\bar g_{33}}{\bar g_{31}^2}\right)^\omega.
\end{array}
\label{eqn.coordmixed}
\end{equation}
Note that they coincide with $\ell^1,\tell^1,\rx^1,\trx^1$ in \eqref{eqn.coordsemi}. The parametrization of the group element is
\begin{equation}
{\scriptsize
g=\begin{pmatrix}
\exp(\bar\chi+uf_1) &
\exp(6\trx+2\phi-\bar\chi-f_1+2f_2+\bar uf_4)&
\exp(\phi+uf_2) \\
-\exp(3\ell-\bar\phi-\bar\chi+uf_3) &
g_{22} &
\exp(3\ell-\phi-\chi+u\bar f_3) \\
\exp(\bar\phi+u\bar f_2) &
-\exp(6\bar\trx+2\bar\phi-\chi-\bar f_1+2\bar f_2+\bar u\bar f_4)&
\exp(\chi+u\bar f_1)
\end{pmatrix},}
\end{equation}
with the Darboux partners given by
\begin{equation}
\begin{array}{l}
\tell  = -\bar f_1+f_2, \\
\rx  = \phi-\bar\chi-2f_1+2f_2.
\end{array}
\end{equation}
On each symplectic leaf, the generating function of the symplectomorphism \eqref{eqn.ylyr} $(\ell,\bar\ell,\tell,\bar\tell)\to(\rx,\bar\rx,\trx,\bar\trx)$ is formally
\begin{equation}
K^{(1,1)} = \int_{\mathcal{O}}^{(\ell,\bar\ell,\trx,\bar\trx)}\theta_1,\qquad \text{where }\theta_1 = \tell\, d\ell+\bar\tell\,d\bar\ell-\rx\,d\trx-\bar\rx\,d\bar\trx,
\label{eqn.11pot}
\end{equation}
where $\mathcal{O}=\mathcal{O}(\phi,\bar\phi,\chi,\bar\chi)$ is collection of base points, one on each symplectic leaf. To find the explicit dependence of $K$ on $\phi,\bar\phi,\chi,\bar\chi$, one has to, as discussed above, solve the second order nonlinear differential equations \eqref{eqn.gerbefp}-\eqref{eqn.gerbefm}. However, this can be circumvented, as discussed in section \ref{sec.genstrat}, by relating the type $(1,1)$ generalized K\"ahler structure to the type $(0,0)$ structure \eqref{eqn.00pot}.

\subsection{T-duality: Relating the two generalized K\"ahler structures}
\label{sec.su3tdual}
\subsubsection{Isometries}
The group of isometries preserving both complex structures of $SU(3)$ is $U(1)_L^2\times U(1)_R^2$, acting on the group element as
\begin{equation}
g\mapsto h_L\,g\,h^{-1}_R\equiv e^{\epsilon\bar h-\bar\epsilon h}g\,e^{\bar\eta\bar h - \eta h},
\label{eqn.su3isom}
\end{equation}
where $h$ and $\bar h$ are defined in \eqref{eqn.algbasis}, and $\epsilon$ and $\eta$ are complex Kac-Moody parameters. In the type $(0,0)$ structure, they obey the constraints
\begin{equation}
\bar {\bD}_+\epsilon=0,~~~~~ {\bD}_\pm\epsilon=0,~~~~~ \bar {\bD}_\pm\eta=0,~~~~~ {\bD}_-\eta=0,
\end{equation}
and the coordinates \eqref{eqn.coordsemi} transform as
\begin{equation}\begin{array}{ll}
\ell^1 \mapsto \ell^1+\eta,~~~~~ &\tell^1\mapsto\tell^1, \\
\ell^2 \mapsto \ell^2+\bar\epsilon+\eta, &\tell^2\mapsto\tell^2+\epsilon+\bar\epsilon+2\eta, \\
\rx^1 \mapsto \rx^1+\eta-\bar\eta, &\trx^1\mapsto\trx^1+\tfrac{1}{2}\bar\eta, \\
\rx^2 \mapsto \rx^2-\eta+\bar\eta, ~~~~~&\trx^2\mapsto\trx^2-\bar\epsilon-\tfrac{1}{2}\bar\eta.
\end{array}\end{equation}
Meanwhile, in the type $(1,1)$ structure, $\epsilon$ and $\eta$ obey (note that, as in the $SU(2)\times U(1)$ case, the chirality constraint on $\epsilon$ differs from that in the type $(0,0)$ structure)
\begin{equation}
\bar {\bD}_\pm\epsilon=0,~~~~~ {\bD}_+\epsilon=0,~~~~~ \bar {\bD}_\pm\eta=0,~~~~~ {\bD}_-\eta=0,
\end{equation}
and the coordinates \eqref{eqn.coordmixed} transform as
\begin{equation}\begin{array}{ll}
\phi \mapsto \phi+\epsilon+\eta,~~~~~&\chi\mapsto\chi+\bar\epsilon+\eta, \\
\ell\mapsto\ell+\eta,&\tell\mapsto\tell, \\
\rx\mapsto\rx+\eta-\bar\eta,&\trx\mapsto\trx+\tfrac{1}{2}\bar\eta.
\end{array}\end{equation}

\subsubsection{T-duality from type $(0,0)$ to type $(1,1)$}
In this subsection, we T-dualize from the type $(0,0)$ structure \eqref{eqn.00pot} to the type $(1,1)$ structure along the left Kac-Moody $U(1)_L$ isometry defined by setting $\epsilon=i\lambda$ and $\eta=0$ in \eqref{eqn.su3isom}, where $\lambda$ is a real parameter. Under this isometry, $\ell^1,\bar\ell^1,\trx^1,\bar\trx^1$ are invariant spectator fields, while $\ell^2\mapsto\ell^2-i\lambda$, $\trx^2\mapsto\trx^2+i\lambda$. The invariance of the Darboux partners $\tell^j,\bar\tell^j,\rx^j,\trx^j$ guarantees that the potential \eqref{eqn.00pot} is invariant this isometry.

The Killing field of the isometry is
\begin{equation}
k = i(-\p_{\ell^2}+\p_{\bar\ell^2}+\p_{\trx^2}-\p_{\bar\trx^2});
\end{equation}
and the invariant combinations $-\ell^2-\bar\ell^2$, $\trx^2+\bar\trx^2$ and $i(\ell^2-\bar\ell^2+\trx^2-\bar\trx^2)$ can be gauged respectively by the components $V^L, V^R$ and $V'$ of the SVM. The T-dual potential is obtained by constraining the SVM to be flat using Lagrange multipliers $\Phi_I$
\begin{equation}
\tilde K(\ell^1,\bar\ell^1,\trx^1,\bar\trx^1,\Phi_I) = K^g(\ell^1,\bar\ell^1,\trx^1,\bar\trx^1,V^I) - V^I\Phi_I,
\label{eqn.dualpot1}
\end{equation}
where $V^I$ are to be eliminated using their equations of motion. We make use of the observation in section \ref{sec.u2obs}: since the T-duality is along a left Kac-Moody isometry, the left complex structure is preserved and continues to be induced by $\bJ_1$, and therefore the relation between the type $(0,0)$ coordinates \eqref{eqn.coordsemi} and type $(1,1)$ coordinates \eqref{eqn.coordphi}, \eqref{eqn.coordchi}, \eqref{eqn.coordmixed} is a solution to the SVM equations of motion. This relation is
\begin{equation}\begin{array}{ll}
\ell^2 &= \chi, \\
\tell^2 &= \phi+\chi, \\
\rx^2 &= \bar\chi-\phi, \\
\trx^2 &= \trx^1 + \phi + f_2.
\end{array}\label{eqn.su3rel}\end{equation}
It is easily verified that this relation is consistent with \eqref{eqn.00y}. The moment maps are linear combinations of the Darboux partners since the isometry acts translationally, so the SVM equations of motion are
\begin{equation}\begin{array}{ll}
\Phi_L &= -\tfrac{1}{2}(\tell^2+\bar\tell^2), \\
\Phi_R &= -\tfrac{1}{2}(\rx^2+\bar\rx^2), \\
\Phi' &= \tfrac{i}{4}(-\tell^2+\bar\tell^2+\rx^2-\bar\rx^2),
\end{array}\end{equation}
which is once again consistent with \eqref{eqn.su3rel}.

As in the $SU(2)\times U(1)$ case, we can write the dual potential as an integral. Note that the invariance of the type $(0,0)$ potential implies
\begin{equation}
\L_kK = i(-\tell^2+\bar\tell^2-\rx^2+\bar\rx^2) = 0,
\end{equation}
which means that the Lagrange multiplier term is
\begin{equation}
V^I\Phi_I = \tell^2(\Phi)\ell^2+\bar\tell^2(\Phi)\ell^2-\rx^2(\Phi)\trx^2-\bar\rx^2(\Phi)\bar\trx^2,
\end{equation}
where $\tell^2,\bar\tell^2,\rx^2,\bar\rx^2$ are understood as functions of $\Phi=(\phi,\bar\phi,\chi,\bar\chi)$ given by \eqref{eqn.su3rel}. Therefore, the dual potential can be written as
\begin{equation}
\tilde K = \int \theta_1 - \ell^2\,d\tell^2 - \bar\ell^2\,d\bar\tell^2 + \trx^2\,d\rx^2 + \bar\trx^2\,d\bar\rx^2.
\end{equation}
In terms of the functions $f_1,f_2$ defined in \eqref{eqn.00param},
\begin{multline}
\tilde K = \int (-\bar f_1+f_2)\, d\ell + 2(f_1-f_2)\, d\trx - (\chi+f_2)\, d\phi + (\bar\phi+\bar f_2)\, d\chi \\
-\frac{1}{2}\left(\phi^2+\chi^2\right)+\trx(\bar\chi-\phi) + \text{cc}.
\label{eqn.11pot2}
\end{multline}
Note that the two terms on the last line are generalized K\"ahler transformations. In contrast to \eqref{eqn.11pot}, this is an unambiguous potential for the type $(1,1)$ structure, obtained without having to solve the nonlinear differential equations \eqref{eqn.gerbefp}-\eqref{eqn.gerbefm}.

\subsubsection{Type $(1,1)$ to type $(0,0)$}
One can also perform the T-duality in the other direction, gauging the same left Kac-Moody isometry of the type $(1,1)$ geometry with an LVM and enforcing it to be flat with semi-chiral Lagrange multipliers, and then integrating out the LVM. This would return \eqref{eqn.11pot2} to the type $(0,0)$ potential \eqref{eqn.00pot}.

\section{Discussion and conclusion}
\setcounter{equation}{0}
\subsection{Results}
In this paper, we studied the generalized K\"ahler structures of two rank 2 groups, $SU(2)\times U(1)$ and $SU(3)$, in detail. We found coordinates that were holomorphic with respect to left-invariant and right-invariant complex structures, and formulae (in one case explicit, in the other implicit) for their generalized K\"ahler potentials (which have an interpretation as the Lagrange density of the sigma model in $(2,2)$ superspace). 

We explained how rank two groups carry generalized K\"ahler structures of two different types, related to one another by T-duality along a Kac-Moody isometry, and how a clever trick trivializes the Legendre transform that is usually needed to relate their generalized K\"ahler potentials. We also gave a wealth of computational details that may be useful for future investigations.

\subsection{Possible future developments}
\subsubsection{Type change}
Apart from $SU(2)\times U(1)$, $U(1)^2$ and their products, left and right complex structures on Lie groups do not commute, and therefore semi-chiral superfields are generically present. Semichiral coordinates are accompanied by type change loci, which are loci of positive codimension on which the type of the generalized geometry changes. For $SU(2)\times U(1)$, an analysis of the type change locus for the type $(0,0)$ generalized geometry was performed in \cite{Sevrin:2011mc}. For $SU(3)$, this has not yet been done and is a direction for future work.

\subsubsection{$(4,4)$ supersymmetry}
The WZW models on both $SU(2)\times U(1)$ and $SU(3)$ admit a further enhancement of supersymmetry to $(4,4)$. Such bi-hypercomplex Lie groups were classified \cite{Spindel:1988sr}, and the whole list of them are $SU(n+1),SU(2n)\times U(1),SO(4n)\times U(1)^{2n},SO(4n+2)\times U(1)^{2n-1},SO(2n+1)\times U(1)^n,Sp(2n)\times U(1)^n, E_6\times U(1)^2,E_7\times U(1)^7,E_8\times U(1)^8,F_4\times U(1)^4,G_2\times U(1)^2$ (and products). There always exists a choice of left and right complex structures that gives rise to a type $(0,0)$ generalized geometry - fully parametrized by semi-chiral coordinates.
For such generalized K\"ahler structures, the potential is an integral of a tautological one-form and can be easily computed without having to solve nonlinear second order PDEs. Generalized geometries of other types on these Lie groups can be obtained from the type $(0,0)$ geometry via T-duality and (possibly repeated) applications of the technique detailed above.

Another direction for the study of $(4,4)$ supersymmetric models is the problem of manifestly realizing all the supersymmetries. For $SU(2)\times U(1)$, a $(4,4)$ formulation of the type $(1,1)$ generalized geometry in biprojective superspace is known \cite{Rocek:1991vk} (although it is not known how to define the contour for the generalized potential), while for the type $(0,0)$ geometry, the extra supersymmetries are not compatible with $(2,2)$ superspace \cite{Lindstrom:2014bra}. Preliminary investigations of the $SU(3)$ model seem to indicate that it is not compatible with the multiplet structure of biprojective superspace. It is worthwhile to conduct a more thorough investigation of off-shell $(4,4)$ supersymmetry for WZW models.

\subsubsection{Other groups}
Another obvious direction is to investigate other groups. The holomorphic structures on $SU(3)$ are surprisingly subtle and we can expect more surprises as we investigate higher rank groups.

\section*{Acknowledgements}
The work of JPA and MR is supported in part by NSF grant PHY1620628. The work of SD and AS is supported in by the ``FWO-Vlaanderen’’ through the project G020714N and an ``aspirant’’ fellowship, and by the Vrije Universiteit Brussel through the Strategic Research Program ``High-Energy Physics’’. We are grateful to the Simons Center for Geometry and Physics for providing a stimulating environment during the conference ``Generalized Geometry and T-dualities’’ where part of this work was performed.

\appendix

\section{Sigma models and supersymmetry}
\setcounter{equation}{0}
\label{sec.appA}
The $d=2$ non-linear sigma model is a Lagrangian field theory of maps $\varphi$ from a two-dimensional worldsheet $(\Sigma,h)$ to a target Riemannian manifold $(M,g)$, equipped with a closed 3-form $H$, known as the torsion. Let $b$ be a local 2-form potential (the Kalb-Ramond 2-form) for $H$, $db=H$. The action is the sum of the integrals of the pullback (via $\varphi$) of $g$, with respect to the volume form of $h$, and the pullback (via $\varphi$) of $b$
\begin{equation}
I[\varphi] = \int_\Sigma \sqrt{h}d^2\sigma\, h^{\alpha\beta}\p_\alpha\varphi^\mu g_{\mu\nu} \p_\beta\varphi^\nu + \int_\Sigma d^2\sigma\, \epsilon^{\alpha\beta}\p_\alpha\varphi^\mu b_{\mu\nu}\p_\beta\varphi^\nu.
\end{equation}
The action depends only on the conformal class of $h$.

Any sigma model admits an $(1,1)$ supersymmetric extension, which moreover can be written in $(1,1)$ superspace as
\begin{equation}
I[\Phi] = \int_\Sigma d^2\sigma\, d^2\theta\, \nabla_+\Phi^\mu(g_{\mu\nu}+b_{\mu\nu})\nabla_-\Phi^\nu
\label{eqn.11act}
\end{equation}
where $\nabla_\pm$ are the $(1,1)$ supercovariant derivatives satisfying the algebra
\begin{equation}
\{\nabla_\pm,\nabla_\pm\}=2i\p_{\stackrel{\doubleplus}{=}},~~~ \{\nabla_\pm,\nabla_\mp\}=0,
\end{equation}
and $\Phi^\mu(\sigma,\theta)$ is the $(1,1)$ superfield which has $\varphi^\mu$ as its bottom component.

To look for extended supersymmetries, one considers the most general transformations
\begin{equation}
\delta\Phi^\mu = \epsilon^+_{(A)}(J_+^{(A)})^\mu{}_\nu \nabla_+\Phi^\nu+\epsilon^-_{(\tilde A)}(J_-^{(\tilde A)})^\mu{}_\nu\nabla_-\Phi^\nu,
\label{eqn.susyansatz}
\end{equation}
where $A=2,\ldots,\N_+$ and $\tilde A=2,\ldots,\N_-$ indexes the extended supersymmetries. Demanding that these transformations satisfy the supersymmetry algebra implies that the $J_+^{(A)}$ and $J_-^{(\tilde A)}$ are integrable complex structures on $M$, and moreover satisfy the Clifford-like relations
\begin{equation}
\{J_+^{(A)},J_+^{(B)}\}=-2\delta^{AB}\mathbbm{1},~~~\{J_-^{(\tilde A)},J_-^{(\tilde B)}\}=-2\delta^{\tilde A\tilde B}\mathbbm{1}.
\end{equation}
Demanding that the action \eqref{eqn.11act} is invariant under these transformations further implies that the metric $g$ is hermitian with respect to all the complex structures $J_\pm^{(A)}$, and that the complex structures are covariantly constant
\begin{equation}
\nabla^{(+)}J_+^{(A)} = 0,~~~ \nabla^{(-)}J_-^{(\tilde A)}=0
\end{equation}
with respect to the Bismut connections $\nabla^{(\pm)}$, which are metric connetions with torsions $\pm g^{-1}H$. Explicitly, the connection coefficients of the Bismut connections are
\begin{equation}
{\Gamma^{(\pm)}}^\mu_{\nu\rho}={\Gamma^{(0)}}^\mu_{\nu\rho}\pm \frac{1}{2}g^{\mu\sigma}H_{\sigma\nu\rho},
\end{equation}
where $\Gamma^{(0)}$ is the Levi-Civita connection.

For $(2,2)$ supersymmetry, which is the subject of discussion in this paper, this implies that the target geometry is bihermitian \cite{Gates:1984nk}, which is equivalent to generalized K\"ahler geometry \cite{Gualtieri:2003dx}. In the more specific case $J_+=\pm J_-$, the manifold is K\"ahler \cite{Zumino:1979et}.

For $(4,4)$ supersymmetry, the target geometry is bi-hypercomplex (or bi-hyperK\"ahler with torsion). The $SU(2)\times U(1)$ and $SU(3)$ WZW models are in fact bi-hypercomplex \cite{Spindel:1988sr}, and therefore has $(4,4)$ supersymmetry, but the off-shell formulation of the supersymmetry is more challenging and will not be addressed in this paper.

\section{Local description of bihermitian geometry}
\setcounter{equation}{0}
\label{sec.gkp}
The bihermitian data $(g,H,J_+,J_-)$ of the generalized K\"ahler manifold may be expressed in terms of generalized K\"ahler potential $K$. Consider a $(2,2)$ sigma model with $N_c$ chiral, $N_t$ twisted chiral and $N_s$ sets of semi-chiral superfields, which locally describes a type $(N_c,N_t)$ generalized K\"ahler manifold of real dimension $2N_c+2N_t+4N_s$. The generalized K\"ahler potential, which also serves as the $(2,2)$ superspace Lagrange density, is a real function of the superfields
\begin{equation}
K = K(\ell,\bar\ell ,\trx,\bar\trx ,\phi,\bar\phi,\chi,\bar\chi).
\end{equation}
We use a convention where $c,\bar c=1,\ldots,N_c$ label the chiral and antichiral superfields, $t,\bar t=1,\ldots,N_t$ label the twisted chiral and twisted antichiral superfields, and $l,\bar l,r,\bar r=1,\ldots,N_s$ label the left, anti-left, right and anti-right semi-chiral superfields respectively. Furthermore, capital indices label the collective set of chiral, twisted chiral and semi-chiral superfields
\begin{equation}
L=(l,\bar l),~~R=(r,\bar r),~~C=(c,\bar c),~~T=(t,\bar t).
\end{equation}
To express the bihermitian data in terms of the potential, we introduce the notation
\begin{equation}
K_{AB} = \begin{pmatrix}K_{ab}&K_{a\bar b}\\ K_{\bar ab}&K_{\bar a\bar b}\end{pmatrix},
\end{equation}
where $A,B=L,R,C,T$, and $K_{ab}$ is shorthand for the second derivative $\p_a\p_bK$. For example, $K_{CL}$ is the $2N_c\times 2N_s$ matrix of second derivatives
\begin{equation}
K_{CL}=\begin{pmatrix}\p_c\p_lK & \p_c\p_{\bar l}K \\ \p_{\bar c}\p_lK & \p_{\bar c}\p_{\bar l}K \end{pmatrix}.
\end{equation}
We write $K_{AB}^{-1}=(K_{BA})^{-1}$. 
We also define
\begin{equation}
C_{AB}=JK_{AB}-K_{AB}J,~~~~~ A_{AB}=JK_{AB}+K_{AB}J,
\end{equation}
where $J$ is the square matrix
\begin{equation}
J=\begin{pmatrix}i\mathbbm{1}&0\\0&-i\mathbbm{1}\end{pmatrix},
\end{equation}
whose size varies depending on the context.

By reducing the $(2,2)$ superspace formulation of the sigma model to $(1,1)$ superspace and eliminating auxiliary fields arising from the semi-chiral sector, one can obtain explicit expressions for the bihermitian data in terms of the potential $K$ (for more details, see \cite{Lindstrom:2005zr}). The complex structures are (in the order $L,R,C,T$)
\begin{align}
J_+ &= \begin{pmatrix} J &&&\\ K_{RL}^{-1}C_{LL} & K_{RL}^{-1}JK_{LR} & K_{RL}^{-1}C_{LC} & K_{RL}^{-1}C_{LT} \\ &&J& \\ &&&J \end{pmatrix}, \nn\\
J_- &= \begin{pmatrix} K_{LR}^{-1}JK_{RL} & K_{LR}^{-1}C_{RR} & K_{LR}^{-1}C_{RC} & K_{LR}^{-1}A_{RT} \\ &J&& \\ &&J& \\ &&&-J \end{pmatrix},
\end{align}
and the Poisson structure $\sigma$ is
\begin{equation}
\sigma = \begin{pmatrix} 0&K_{LR}^{-1}&0&0 \\ -K_{RL}^{-1}&0&0&0\\ 0&0&0&0\\0&0&0&0 \end{pmatrix}.
\end{equation}
Note that a change of coordinates from $(\ell,\bar\ell ,\trx,\bar\trx ,\phi,\bar\phi,\chi,\bar\chi)$ to $(\ell,\bar\ell,\tell,\bar\tell,\phi,\bar\phi,\chi,\bar\chi)$, where
\begin{equation}
\tell  := \frac{\p K}{\p \ell},~~~ \bar\tell :=\frac{\p K}{\p \bar\ell }
\label{eqn.yl1}
\end{equation}
diagonalizes $J_+$ and puts the Poisson structure in the canonical form
\begin{equation}\sigma=\displaystyle
\begin{pmatrix}0&\mathbbm{1}&0&0\\-\mathbbm{1}&0&0&0\\0&0&0&0\\0&0&0&0\end{pmatrix};
\end{equation}
while a coordinate change to $(\rx,\bar\rx ,\trx ,\bar\trx ,\phi,\bar\phi,\chi,\bar\chi)$, where
\begin{equation}
\rx  := -\frac{\p K}{\p\trx},~~~ \bar\rx :=-\frac{\p K}{\p\bar\trx}
\label{eqn.yr}
\end{equation}
diagonalizes $J_-$ and again puts $\sigma$ in the canonical form. In other words, the generalized potential $K$ serves as the generating function of the symplectomorphism between the $J_+$-holomorphic Darboux coordinates $(\ell,\bar\ell ,\tell ,\bar\tell )$ and $J_-$-holomorphic Darboux coordinates $(\rx,\bar\rx ,\trx ,\bar\trx )$ on the symplectic leaves of $\sigma$. This is an important characterization of the generalized K\"ahler potential which will be repeatedly used in this paper.

The metric $g$ and 2-form potential $b$ can be obtained from the local symplectic forms
\begin{align}
\F_+ &= \frac{1}{2}(b_+-g)J_+ =
\frac{1}{2}\begin{pmatrix} d\ell^L \\ d\rx^R \\ d\phi^C \\ d\chi^T \end{pmatrix}^T
\begin{pmatrix} -C_{LL} & -A_{LR} & -A_{LC} & -C_{LT} \\ 
A_{RL} & C_{RR} & C_{RC} & A_{RT} \\
A_{CL} & C_{CR} & C_{CC} & A_{CT} \\
-C_{TL} & -A_{TR} & -A_{TC} & -C_{TT} \end{pmatrix}
\begin{pmatrix} d\ell^L \\ d\rx^R \\ d\phi^C \\ d\chi^T \end{pmatrix}\, \label{eqn.gerbefp} \\
\F_- &= \frac{1}{2}(b_-+g)J_- =
-\frac{1}{2}\begin{pmatrix} d\ell^L \\ d\rx^R \\ d\phi^C \\ d\chi^T \end{pmatrix}^T
\begin{pmatrix} C_{LL} & C_{LR} & C_{LC} & C_{LT} \\
C_{RL} & C_{RR} & C_{RC} & C_{RT} \\
C_{CL} & C_{CR} & C_{CC} & C_{CT} \\
C_{TL} & C_{TR} & C_{TC} & C_{TT} \end{pmatrix}
\begin{pmatrix} d\ell^L \\ d\rx^R \\ d\phi^C \\ d\chi^T \end{pmatrix}. \label{eqn.gerbefm}
\end{align}
Here, $b_+$ and $b_-$ are different two-form potentials for $H$, $db_\pm=H$, chosen such that $b_\pm$ are $(2,0)+(0,2)$ forms with respect to $J_\pm$. $b_+$ and $b_-$ differ by an exact form
\begin{equation}
\G = \frac{1}{2}(b_+-b_-) = d\lambda.
\end{equation}
The local symplectic two-forms $\F_\pm$ may be interpreted as connections for flat gerbes \cite{Hull:2008vw}.

Finally, we turn to the quantum properties of the underlying $(2,2)$ sigma model. Denoting
\begin{equation}
N_+ = \begin{pmatrix} K_{l\bar l} & K_{lr} & K_{l\bar t} \\ K_{\bar r\bar l} & K_{\bar rr} & K_{\bar r\bar t} \\ K_{t\bar l} & K_{tr} & K_{t\bar t} \end{pmatrix},~~~~~ N_- = \begin{pmatrix} K_{l\bar l} & K_{l\bar r} & K_{l\bar c} \\ K_{r\bar l} & K_{r\bar r} & K_{r\bar c} \\ K_{c\bar l} & K_{c\bar r} & K_{c\bar c} \end{pmatrix}, \end{equation}
the one loop beta function vanishes if \cite{Grisaru:1997pg}
\begin{equation}
\log\frac{\det N_+}{\det N_-} = f(\ell,\phi,\chi) + \bar f(\bar\ell,\bar\phi,\bar\chi) + g(\rx,\phi,\bar\chi) + \bar g(\bar\trx,\bar\phi,\chi),
\end{equation}
while the target manifold is generalized Calabi-Yau if it satisfies the stronger condition \cite{Hull:2010sn}
\begin{equation}
\log\frac{\det N_+}{\det N_-} = \text{const}.
\end{equation}

We end this section with a final remark. Mapping the bihermitian data,
\begin{eqnarray}
(g,H,J_+,J_-) \rightarrow (g,H,J_+,-J_-)\,,
\end{eqnarray} 
merely amounts to mapping the corresponding generalized K\"ahler potential 
\begin{eqnarray}
K(\ell,\bar\ell ,\trx,\bar\trx ,\phi,\bar\phi,\chi,\bar\chi) \rightarrow 
-K(\ell,\bar\ell ,\bar\trx,\trx ,\chi,\bar\chi ,\phi,\bar\phi)\,,
\end{eqnarray} 
leaving any of the expressions above unchanged. This is a general symmetry of (2,2) supersymmetric sigma models.

\section{Isometries and T-duality}
\setcounter{equation}{0}
\label{sec.tduality}
T-duality is an equivalence of the underlying physics of sigma models describing different geometries. The duality can be realized by gauging an isometry of the sigma model, and then adding a Lagrange multiplier enforcing flatness of the gauge connection, so that it is equivalent to the original model. Integrating out instead the non-dynamical gauge connection yields the T-dual model, which in general describes a different geometry \cite{Rocek:1991ps}.

In $(2,2)$ superspace, T-duality also changes the complex structures \cite{Ivanov:1994ec}, and therefore also the type of the generalized K\"ahler geometry. Isometries of $(2,2)$ supersymmetric sigma models fall into three categories: in appropriate local coordinates, they act either on (i) a chiral or a twisted chiral coordinate, (ii) a pair of chiral and twisted chiral coordinates, or (iii) a set of semi-chiral coordinates.\footnote{More generally, isometries of type (i) and (ii) can also act on semi-chiral coordinates.} The isometries can then be gauged using an appropriate gauge connection: for (i), a usual vector multiplet; for (ii), a large vector multiplet (LVM); and for (iii) a semi-chiral vector multiplet (SVM) \cite{Lindstrom:2007vc,Lindstrom:2007sq,Lindstrom:2008hx}.

T-duality along isometries of type (i) exchanges a chiral superfield for a twisted chiral (or vice versa), so if the original generalized geometry is of type $(p,q)$, then the dual geometry has type $(p-1,q+1)$ (or $(p+1,q-1)$). T-dualizing along a type (ii) isometry exchanges a pair of chiral and twisted chiral coordinates for a set of semi-chirals, so the dual model has type $(p-1,q-1)$. Finally, for type (iii), a set of semi-chirals is exchanged for a chiral and twisted chiral, so the dual model has type $(p+1,q+1)$.

The LVM and SVM are novel vector multiplets which do not arise in the K\"ahler (torsionless) case. We briefly review these vector multiplets.

\subsection{Large vector multiplet (LVM)}
A type (i) isometry $k=k^\phi\p_\phi+k^{\bar\phi}\p_{\bar\phi}+k^\chi\p_\chi+k^{\bar\chi}\p_{\bar\chi}$ acts on both chiral $\phi$ and twisted chiral $\chi$ coordinates and is gauged with a large vector multiplet (LVM), consisting of three real vector multiplets $V^I=(V^\phi,V^\chi,V')$. Since the isometry preserves the generalized K\"ahler structure $\L_kg=\L_kH=\L_kJ_\pm=0$, the potential transforms into a generalized K\"ahler transformation
\begin{equation}
\mathcal{L}_kK=f(\phi,\chi)+\bar f(\bar\phi,\bar\chi)+g(\phi,\bar\chi)+\bar g(\bar\phi,\chi).
\end{equation}
The preservation of both complex structures $J_\pm$ implies that the components of $k$ depend only on coordinates of the appropriate chirality
\begin{equation}
k^\phi=k^\phi(\phi),~~~ k^{\bar\phi}=k^{\bar\phi}(\bar\phi),~~~ k^\chi=k^\chi(\chi),~~~ k^{\bar\chi}=k^{\bar\chi}(\bar\chi).
\end{equation}
In this paper we deal only with invariant potentials, so we shall assume $f=\bar f=g=\bar g=0$. Therefore,
\begin{equation}\begin{array}{ll}
k^\phi\p_\phi K =& i\mu_\phi+\mu', \\
k^{\bar\phi}\p_{\bar\phi}K =& -i\mu_\phi+\mu', \\
k^\chi\p_\chi K =& i\mu_\chi-\mu', \\
k^{\bar\chi}\p_{\bar\chi}K =& -i\mu_\chi-\mu',
\end{array}\end{equation}
for some real functions $\mu_I=(\mu_\phi,\mu_\chi,\mu')$. These functions have the interpretation as (local) moment maps with respect to the 2-form gerbe potentials $\mathcal{F}_\pm$ of the generalized geometry, and another symplectic form $\mathcal{G}$ we define below. Recall that generalized K\"ahler geometry may be reformulated as a flat biholomorphic gerbe \cite{Hull:2008vw} with local two-form potentials
\begin{equation}
\mathcal{F}_\pm = \frac{1}{2}(b_\pm J_\pm\mp\omega_\pm) = \frac{i}{2}d(\mp K_\phi d\phi\pm K_{\bar\phi}d\bar\phi+K_\chi d\chi-K_{\bar\chi}d\bar\chi) = \mp\frac{1}{2}dd_\mp^cK,
\end{equation}
where subscripts of $K$ denote derivatives, and $b_\pm$ are local torsion potentials $db_\pm=H$ which are chosen to be $(2,0)+(0,2)$ with respect to the complex structures $J_\pm$. We also introduce the local symplectic form
\begin{equation}
\mathcal{G}=\frac{1}{2}(b_+-b_-)=\frac{1}{2}d(-K_\phi d\phi-K_{\bar\phi}d\bar\phi+K_\chi d\chi+K_{\bar\chi}d\bar\chi)=\frac{1}{2}d(K_\mu\, J_+J_-dX^\mu).
\end{equation}
For an invariant potential $K$,
\begin{align}
\iota_k \mathcal{F}_\pm=& \mp\frac{1}{2}\mathcal{L}_k(K_\mu\, J_\mp dX^\mu)\pm\frac{1}{2}d\iota_k(K_\mu\, J_\mp dX^\mu)\nonumber \\
=& -\frac{1}{2}d(\pm\mu_\phi-\mu_\chi),
\end{align}
\begin{align}
\iota_k \mathcal{G} =& \frac{1}{2}\mathcal{L}_k(K_\mu\, J_+J_-dX^\mu)-\frac{1}{2}d\iota_k(K_\mu\, J_+J_-dX^\mu) = d\mu',
\end{align}
which shows that the $\mu_I$ are indeed moment maps with respect to $\F_\pm$ and $\G$.

Gauging the isometry in $(2,2)$ superspace promotes the parameter to chiral and twisted chiral parameters
\begin{align}
\delta_\Lambda =& \Lambda^\phi k^\phi\p_\phi+\Lambda^{\bar\phi}k^{\bar\phi}\p_{\bar\phi}+\Lambda^\chi k^\chi\p_\chi+\Lambda^{\bar\chi}k^{\bar\chi}\p_{\bar\chi} \nonumber \\
=& \frac{1}{4}(\Lambda^\phi+\Lambda^{\bar\phi}+\Lambda^\chi+\Lambda^{\bar\chi})\mathcal{L}_k+i(\Lambda^\phi-\Lambda^{\bar\phi})\mathcal{L}_{k_{(\phi)}}+i(\Lambda^\chi-\Lambda^{\bar\chi})\mathcal{L}_{k_{(\chi)}}+(\Lambda^\phi+\Lambda^{\bar\phi}-\Lambda^\chi-\Lambda^{\bar\chi})\mathcal{L}_{k_{(`)}}.
\end{align}
where the complex conjugate vector fields $k_I$ are defined by
\begin{equation}\begin{array}{lll}
k_{(\phi)}=&\displaystyle-\frac{1}{4}(J_++J_-)k &= \displaystyle\frac{i}{2}(k^{\bar\phi}\p_{\bar\phi}-k^\phi\p_\phi), \\[5pt]
k_{(\chi)}=&\displaystyle-\frac{1}{4}(J_+-J_-)k &=\displaystyle \frac{i}{2}(k^{\bar\chi}\p_{\bar\chi}-k^\chi\p_\chi), \\[5pt]
k_{(')}=& \displaystyle-\frac{1}{4}J_+J_-k &= \displaystyle\frac{1}{4}(k^\phi\p_\phi+k^{\bar\phi}\p_{\bar\phi}-k^\chi\p_\chi-k^{\bar\chi}\p_{\bar\chi}).
\end{array}\end{equation}
The gauged potential is
\begin{align}
K^g(\phi,\bar\phi,\chi,\bar\chi,V^\phi,V^\chi,V')
=& K(\phi,\bar\phi,\chi,\bar\chi) + \int_0^1 dt\ \exp(tV^I\mathcal{L}_{k_I})V^K\mu_K \nn\\
=& \exp(V^I\mathcal{L}_{k_I})K(\phi,\bar\phi,\chi,\bar\chi),
\label{eqn.kg2}
\end{align}
with implicit sums over repeated indices $I,J,K$ (over $\phi,\chi,'$). Note that $\L_{k_I}K=\mu_I$.

The equations of motion of $V^I$ from \eqref{eqn.kg2} are
\begin{equation}
\frac{\p K^g}{\p V^I}=\exp(V^J\mathcal{L}_{k_J})\mathcal{L}_{k_I}K=\exp(V^J\mathcal{L}_{k_J})\mu_I.
\end{equation}
Since
\begin{equation}
\delta_\Lambda K^g = \exp(V^J\mathcal{L}_{k_J})\left(\delta_\Lambda K+\mathcal{L}_{k_I}K\, \delta_\Lambda V^I\right),
\end{equation}
gauge invariance of \eqref{eqn.kg2} follows provided
\begin{equation}
\begin{array}{l}
\delta_\Lambda V^\phi = i(\Lambda^{\bar\phi}-\Lambda^{\phi}), \\
\delta_\Lambda V^\chi = i(\Lambda^{\bar\chi}-\Lambda^\chi), \\
\delta_\Lambda V' = -\Lambda^\phi-\Lambda^{\bar\phi}+\Lambda^\chi+\Lambda^{\bar\chi}.
\end{array}
\end{equation}

It is convenient to combine the gauge fields into complex combinations
\begin{equation}
\mathbb{V}=\frac{1}{2}(-V'+i(V^\phi-V^\chi)),\qquad \tilde{\mathbb{V}}=\frac{1}{2}(-V'+i(V^\phi+V^\chi)),
\end{equation}
which transform with semi-chiral parameters $\delta_\Lambda\mathbb{V}=\Lambda^\phi-\Lambda^\chi,\delta_\Lambda\tilde{\mathbb{V}}=\Lambda^\phi-\Lambda^{\bar\chi}$. Therefore, the following are four gauge invariant semi-chiral field strengths
\begin{equation}
\begin{array}{l}
\mathbb{G}_+=\bar {\bD}_+\mathbb{V},~~~\bar\bG_+={\bD}_+\bV, \\
\mathbb{G}_-=\bar {\bD}_-\tilde{\mathbb{V}},~~~ \bar\bG_-={\bD}_-\tilde\bV.
\end{array}
\end{equation}

To enforce the flatness of the LVM, one constrains its field strengths with Lagrange multipliers of the appropriate semi-chirality,
\begin{align} K_{\text{LM}} =& -\frac{1}{2}V^IX_I = -(\ell\mathbb{V}+\rx\tilde{\mathbb{V}}+\bar \ell\bar{\mathbb{V}}+\bar \rx\bar{\tilde{\mathbb{V}}}) \nn\\
=& -\frac{1}{2}(V'X'+V^\phi X_\phi+V^\chi X_\chi),
\end{align}
where $X'=-(\ell+\bar\ell +\rx+\bar\rx ),X_\phi=i(\ell-\bar\ell +\rx-\bar\rx ),X_\chi=i(-\ell+\bar\ell +\rx-\bar\rx )$. Here $I$ runs over the components $\phi,\chi,'$ of the vector multiplet.

To obtain the T-dual sigma model, one eliminates the flat vector fields $V^I$ by their equations of motion, which set the moment maps $\mu_I$ equal to the Fayet-Iliopolous terms $X_I$.

\subsection{Semichiral vector multiplet (SVM)}
A type (ii) isometry $k=k^\ell\p_\ell+k^{\bar\ell}\p_{\bar\ell}+k^\rx\p_\rx+k^{\bar\rx}\p_{\bar\rx}$ acting on semi-chiral coordinates $\ell,\bar\ell,\rx,\bar\rx$ is gauged with a semi-chiral vector multiplet (SVM) consisting of three real vector multiplets $V^I=(V^L,V^R,V')$. As discussed in Section \ref{sec.22nlsm}, the choice of coordinates $\ell,\bar\ell,\rx,\bar\rx$ encodes a choice of polarization on the symplectic leaves of the generalized K\"ahler manifold; we demand that the isometry $k$ preserves this polarization (on top of the usual conditions of a generalized K\"ahler isometry). This implies
\begin{align}
&\mathcal{L}_kK = f(\ell)+\bar f(\bar\ell)+g(\rx)+\bar g(\bar\rx), \label{eqn.isom3} \\
&k^\ell=k^\ell(\ell),~~~ k^\rx=k^\rx(\rx),~~~ k^{\bar\ell}=k^{\bar\ell}(\bar\ell),~~~ k^{\bar\rx}=k^{\bar\rx}(\bar\rx).
\end{align}
For the cases encountered in this paper, the potential $K$ is invariant, so we shall set $f=\bar f=g=\bar g=0$. Note that the mixed coordinate system $(\ell,\bar\ell,\rx,\bar\rx)$ is not holomorphic with respect to either complex structure, so that e.g. $k^\ell$ may not be $J_+$ holomorphic despite it depending only on $\ell$.

From \eqref{eqn.isom3} it follows that
\begin{equation}\begin{array}{ll}
k^\ell\p_\ell K =& i\mu_L+\mu', \\
k^{\bar\ell}\p_{\bar\ell}K =& -i\mu_L+\mu', \\
k^\rx\p_\rx K =& i\mu_R-\mu', \\
k^{\bar\rx}\p_{\bar\rx}K =& -i\mu_R-\mu',
\end{array}\end{equation}
where $\mu_I=(\mu_L,\mu_R,\mu')$ are three real functions. In fact, the $\mu_I$ may be interpreted as moment maps. Recall that on a symplectic leaf of a generalized K\"ahler manifold, the inverse of the Poisson structure $\pi$ is a symplectic form $\Omega=g[J_+,J_-]^{-1}$ which may be thought of as the real part of a holomorphic symplectic form with respect to \emph{either} complex structure, $\Omega=\Omega_L+\bar\Omega_L=\Omega_R+\bar\Omega_R$, where $\Omega_L=-d(\p_\ell K\, d\ell),\Omega_R=-d(\p_\rx K\, d\rx)$. Then, using subscripts to denote derivatives of $K$, we may compute
\begin{align}
\iota_k\Omega_L =& k^\ell dK_\ell - (k^{\bar\ell}K_{\ell\bar\ell}+k^\rx K_{\ell\rx}+k^{\bar\rx}K_{\ell\bar\rx})d\ell \nonumber\\
=& d(k^\ell K_\ell)-\p_\ell(k^\ell K_\ell+k^{\bar\ell}K_{\bar\ell}+k^\rx K_\rx+k^{\bar\rx}K_{\bar\rx})d\ell \nonumber\\
=& d(k^\ell K_\ell) = d(i\mu_L+\mu'),
\end{align}
and similarly for $\Omega_R$. This shows that $(\mu^L,\mu^R,\mu')$ are moment maps for the symplectic forms $\im(\Omega_L),\im(\Omega_R)$ and $\re(\Omega_L)=\re(\Omega_L)$ respectively.

Upon gauging, each $J_\pm$-(anti)holomorphic and polarized sector acquires its own gauge parameter, $\Lambda^L,\bar\Lambda^{L},\Lambda^R,\bar\Lambda^{R}$, so that the gauge variation is
\begin{align}
\delta_\Lambda =& \Lambda^Lk^\ell\p_\ell + \bar\Lambda^{L}k^{\bar\ell}\p_{\bar\ell} + \Lambda^Rk^\rx\p_\rx + \bar\Lambda^{R}k^{\bar\rx}\p_{\bar\rx} \nonumber\\
=& \frac{1}{4}(\Lambda^L+\bar\Lambda^{L}+\Lambda^R+\bar\Lambda^{R})\mathcal{L}_k+i(\Lambda^L-\bar\Lambda^{L})\mathcal{L}_{k_{L}}+i(\Lambda^R-\bar\Lambda^{R})\mathcal{L}_{k_{R}}+(\Lambda^L+\bar\Lambda^{L}-\Lambda^R-\bar\Lambda^{R})\mathcal{L}_{k'},
\label{eqn.lambda1}
\end{align}
where
\begin{equation}\begin{array}{ll}
k_{L} =& \displaystyle\frac{i}{2}(k^{\bar\ell}\p_{\bar\ell}-k^\ell\p_\ell) \\[5pt]
k_{R} =& \displaystyle\frac{i}{2}(k^{\bar\rx}\p_{\bar\rx}-k^\rx\p_\rx) \\[5pt]
k'=& \displaystyle\frac{1}{4}(k^\ell\p_\ell+k^{\bar\ell}\p_{\bar\ell}-k^\rx\p_\rx-k^{\bar\rx}\p_{\bar\rx}).
\end{array}\end{equation}
Note that $\mathcal{L}_{k_I}K=\mu_I$. The gauged action is
\begin{align}
K^g(\ell,\bar\ell,\rx,\bar\rx,V^L,V^R,V')
=& K(\ell,\bar\ell,\rx,\bar\rx) + \int_0^1 \exp(tV^I\mathcal{L}_{k_I})(V^K\mu_K) \nn\\
=& \exp(V^I\mathcal{L}_{k_I})K.
\label{eqn.kg3}
\end{align}

The equations of motion of $V^I$ from \eqref{eqn.kg3} are
\begin{equation}
\frac{\p K^g}{\p V^I} = \exp(V^J\mathcal{L}_{k_J})\mathcal{L}_{k_I}K
= \exp({V^J\mathcal{L}_{k_J}})\mu_I.
\label{eqn.svmaux}
\end{equation}
The gauge variation of $K^g$ is
\begin{equation}
\delta_\Lambda K^g = \exp(V^J\mathcal{L}_{k_J})\left(\delta_\Lambda K + \mathcal{L}_{k_I}K\, \delta_\Lambda V^I\right).
\end{equation}
Using \eqref{eqn.lambda1} and $\mathcal{L}_kK=0$, we see that $\delta_\Lambda K^g$ vanishes provided the SVM transforms as
\begin{equation}
\begin{array}{ll}
\delta_\Lambda V^L =& i(\bar\Lambda^{L}-\Lambda^L), \\
\delta_\Lambda V^R =& i(\bar\Lambda^{R}-\Lambda^R), \\
\delta_\Lambda V' =& -\Lambda^L-\bar\Lambda^{L}+\Lambda^R+\bar\Lambda^{R}.
\end{array}
\end{equation}

The gauge invariant field strengths are built out of the complex combinations
\begin{equation}
\mathbb{V}=\frac{1}{2}(-V'+i(V^L-V^R)),~~~\tilde{\mathbb{V}}=\frac{1}{2}(-V'+i(V^L+V^R))
\end{equation}
which transform as $\delta_\Lambda\mathbb{V}=\Lambda^L-\Lambda^R$, $\delta_\Lambda\tilde{\mathbb{V}}=\Lambda^L-\bar\Lambda^R$. The complex field strengths are
\begin{equation}\begin{array}{l}
\mathbb{F}=\bar {\bD}_+\bar {\bD}_-\mathbb{V},~~~\bar\bF={\bD}_+{\bD}_-\bar\bV, \\
\tilde\bF=\bar {\bD}_+{\bD}_-\tilde\bV,~~~\bar{\tilde\bF}={\bD}_+\bar {\bD}_-\bar{\tilde\bV},
\end{array}\end{equation}
which are respectively chiral and twisted chiral.

To enforce the flatness of the SVM, one constrains its field strengths with Lagrange multipliers of the appropriate chirality,
\begin{align}
K_{\text{LM}} =& -\frac{1}{2}V^I\Phi_I = -(\phi\mathbb{V}+\bar\phi\bar{\mathbb{V}}+\chi\tilde{\mathbb{V}}+\bar\chi\bar{\tilde{\mathbb{V}}})\nn \\
=& -\frac{1}{2}(V'\Phi'+V^L\Phi_L+V^R\Phi_R),
\end{align}
where $\Phi'=-(\phi+\bar\phi+\chi+\bar\chi),\Phi_L=i(\phi-\bar\phi+\chi-\bar\chi),\Phi_R=i(-\phi+\bar\phi+\chi-\bar\chi)$.

To obtain the T-dual sigma model, one eliminates the flat vector fields $V^I$ by their equations of motion, which set the moment maps $\mu_I$ equal to the Fayet-Iliopolous terms $\Phi_I$.

\section{Finding complex coordinates}
\setcounter{equation}{0}
\label{app.holcoord}
Suppose we have a Lie algebra complex structure (with $T_\alpha=(T_a,T_{\bar a})$ denoting respectively the holomorphic and antiholomorphic generators) which induces a complex structure on the Lie group, with holomorphic coordinates $x^\mu=(z^i,\bar z^{\bar i})$. For definiteness, suppose we are working with the left complex structure, so the complex structures on the Lie algebra and group are related by conjugation by the left Maurer-Cartan frame $g^{-1}dg=e^\alpha T_\alpha$. In this appendix, we discuss how to obtain the holomorphic coordinates $z^i,\bar z^{\bar i}$ in some neighborhood around the origin from the Lie algebra complex structure.

Compatibility of the complex structures of the group and algebra implies that the Maurer-Cartan frames corresponding to the holomorphic Lie algebra generators lie in the holomorphic cotangent bundle:
\begin{equation}
e_{\bar i}^a d\bar z^{\bar i}=0,~~~~~ e_i^{\bar a}dz^i=0.
\label{eqn.holframe}
\end{equation}
Suppose we parametrize the group element as $g=\exp(\xi^a(z,\bar z)T_a-\bar\xi^{\bar a}(z,\bar z)T_{\bar a})$. In the cases encountered in this paper, the group is unitary, and if $T_a$ are chosen to be hermitian, then $\bar\xi^{\bar a}=(\xi^a)^\ast$. The Maurer-Cartan frames are
\begin{align}
e^\alpha T_\alpha &= \exp(-\L_{\xi-\bar\xi})d = \sum_{n=0}^\infty \frac{1}{(n+1)!}(-\L_{\xi-\bar\xi})^n d(\xi-\bar\xi) \nn\\
&= d(\xi-\bar\xi) + \frac{1}{2}[d(\xi-\bar\xi),\xi-\bar\xi] + \frac{1}{3!}[[d(\xi-\bar\xi),\xi-\bar\xi],\xi-\bar\xi] + \ldots ,
\label{eqn.mcframe}
\end{align}
where $\L_X(Y)=[X,Y]$.
In a neighborhood around the origin, we can solve for $\xi(z,\bar z)$ order by order, by imposing \eqref{eqn.holframe}. Suppose that $z=\bar z=0$ at the origin, and expand $\xi$ around it
\begin{equation}
\xi^\alpha(z,\bar z) = A^\alpha_i z^i + A^\alpha_{\bar i}\bar z^{\bar i} + A^\alpha_{ij}z^iz^j + A^\alpha_{\bar ij}\bar z^{\bar i}z^j + A^\alpha_{\bar i\bar j}\bar z^{\bar i}\bar z^{\bar j} + \ldots
\end{equation}
Using a holomorphic diffeomorphism, we can set $A^a_i=\delta_i^a$ and $A^a_{ij\ldots k}=0$. This allows us to identify the holomorphic coordinates $z^i$ with the holomorphic directions at the origin $e^a|_{g=1}$ determined by the Lie algebra complex structure. We substitute this expansion into \eqref{eqn.mcframe} and apply the constraint \eqref{eqn.holframe}, order by order in $z$ and $\bar z$. At leading order, we obtain
\begin{equation}
A^a_{\bar i}=0,~~~~~ A^{\bar a}_i=0,
\end{equation}
and the next order yields
\begin{equation}
A^a_{\bar i\bar j}=0,~~~~~ A^a_{\bar ij}=\frac{1}{2}f_{\bar ij}^a,~~~~~ A^{\bar a}_{ij}=0,~~~~~ A^{\bar a}_{\bar ij}=\frac{1}{2}f_{\bar ij}^{\bar a},
\end{equation}
where the integrability condition (see section \ref{sec.22wzw}) $f_{bc}^{\bar a}=0,f_{\bar b\bar c}^a=0$ has been used. This process can be iterated, and in principle yields all the coefficients $A^\alpha_{\bar i\bar j\ldots kl}$ in terms of the structure constants. The integrability condition ensures that solutions for the coefficients always exists. This yields a series for $\xi(z,\bar z)$ and therefore a complex parametrization of the group in a neighborhood of the origin within the radius of convergence.

An important check on our expressions for the holomorphic coordinates comes from the integrability condition that the form
\begin{equation}
\Omega=\bigwedge_a\tr(T_a g^{-1}dg),
\label{topform.D}
\end{equation}
is proportional to the holomorphic top form, and hence annihilates all holomorphic differentials $dz^a$. 

\section{Other type $(0,0)$ potentials for $SU(2)\times U(1)$}
\setcounter{equation}{0}
\label{sec.appgkp00}
In section \ref{sec.gkp00}, we found a type $(0,0)$ generalized K\"ahler potential for $SU(2)\times U(1)$ corresponding to a particular choice of parametrization and polarization. Here we explore other choices.

One choice of parametrization is
\begin{equation}
\ell=z_+^2,~~~~~\tell=z_+^1,~~~~~\rx=-z_-^1,~~~~~\trx=z_-^2,
\end{equation}
satisfying $\re d(\tell\ d\ell+\trx\ dr)=0$. In the polarization spanned by $\tell,\bar\tell$ and $\trx,\bar\trx$, the group element is parametrized as
\begin{equation}
g=\begin{pmatrix} e^{\bar\zeta\theta+\trx} & e^{-\zeta\theta+\tell} \\ -e^{-\zeta\theta+\bar\tell} & e^{\bar\zeta\theta+\bar\trx}\end{pmatrix}.
\label{eqn.klr3param}
\end{equation}
Unimodularity of the $SU(2)$ factor implies that $\theta$ satisfies
\begin{equation}
e^{\theta+\trx+\bar\trx}+e^{-\theta+\tell+\bar\tell}=1,
\end{equation}
which is solved by
\begin{equation}
\theta = \tell+\bar\tell+\log f(e^{\tell+\bar\tell+\trx+\bar\trx}),
\end{equation}
where $f(x)$ solves the quadratic equation
\begin{equation}
x\, f^2(x) - f(x) + 1 = 0.
\end{equation}
The potential, satisfying $\frac{\p K}{\p\tell}=-\ell,\frac{\p K}{\p\trx}=-\rx$ is
\begin{equation}
K^{(0,0)}_2 = \tell\trx+\bar\tell\bar\trx+\frac{1}{2}(\trx+\bar\trx)^2+\int^{\tell+\bar\tell+\trx+\bar\trx}dx\ \log f(e^x).
\label{eqn.klr3}
\end{equation}
Note that this parametrization \eqref{eqn.klr3param} is related to the parametrization \eqref{eqn.klr1param} in Section \ref{sec.gkp00} by the symplectomorphism $(\rx,\trx)\mapsto(\rx',\trx')$ with $\rx'=\rx+\trx,\trx'=\trx$, so we expect the potentials \eqref{eqn.klr3} and \eqref{eqn.klr1} to be related by a generalized K\"ahler transformation followed by a change in polarization. Indeed, it can be checked that
\begin{equation}
K^{(0,0)}_0(\tell,\bar\tell,\rx,\bar\rx) = K^{(0,0)}_2(\tell,\bar\tell,\trx,\bar\trx)-\frac{1}{2}(\trx^2+\bar\trx^2)+\rx\trx+\bar\rx\bar\trx.
\end{equation}
This potential \eqref{eqn.klr3} can be obtained via T-duality along $U(1)$ from the type $(1,1)$ potential \eqref{eqn.kbilp1}, provided the following generalized K\"ahler transformation is added
\begin{multline}
K^{(1,1)}_0 - \frac{1}{2}(\phi-\chi)^2 - \frac{1}{2}(\bar\phi-\bar\chi)^2 \\
=-\frac{1}{2}(\phi-\bar\phi)^2+\frac{1}{4}(\chi-\bar\chi-\phi+\bar\phi)^2-\frac{1}{4}(\phi+\bar\phi-\chi-\bar\chi)^2 + \int^{\phi+\bar\phi-\chi-\bar\chi}dq\ \log(1+e^q).
\end{multline}

Another family of parametrizations, indexed by a nonzero real parameter $\gamma$, is given by
\begin{equation}\begin{array}{ll}
\ell = \displaystyle\frac{1}{2}(\gamma z_+^1-(\gamma+1)z_+^2),~~~~~ &\tell = 2(-z_+^1+z_+^2), \\[5pt]
\rx = \ds\frac{1}{2}(-\gamma z_-^1+(\gamma-1)z_-^2),~~~~~ &\trx = 2(-z_-^1+z_-^2).
\end{array}\end{equation}
In the polarization $\ell,\bar\ell$ and $\rx,\bar\rx$, the potential is
\begin{equation}
K^{(0,0)}_3 = \frac{1}{2\gamma^2}(\ell-\bar\ell+\rx-\bar\rx)^2+\frac{1}{\gamma}\left((\rx-\bar\rx)^2-(\ell-\bar\ell)^2\right) - \int^{\ell+\bar\ell+\rx+\bar\rx}dx\ \log(e^x-1).
\end{equation}
This potential can be obtained via T-duality along $U(1)$ from the type $(1,1)$ potential \eqref{eqn.kbilp1}, with the addition of the generalized K\"ahler transformation
\begin{multline}
K^{(1,1)}_0 + \frac{\gamma}{2}\left((\phi-\chi)^2 - (\phi-\bar\chi)^2 + (\bar\phi-\bar\chi)^2 - (\bar\phi-\chi)^2\right) \\
= \frac{1}{2}(\chi-\bar\chi)^2 - \gamma(\chi-\bar\chi)(\phi-\bar\phi) + \int^{\phi+\bar\phi-\chi-\bar\chi}dq\ \log(1+e^q).
\end{multline}

\section{T-duality from type $(1,1)$ to type $(0,0)$ on $SU(2) \times U(1)$}
\label{sec.su2u1tdual}
We begin with the type $(1,1)$ generalized K\"ahler structure, with group element parametrized as in \eqref{eqn.bilpparam}. We may perform the T-duality along any factor of the $U(1)_L\times U(1)_R\times U(1)$ Kac-Moody isometry group. The complex structures $J_\pm$ map these isometries into one another, so in superspace, where the gauge group is complexified, the gauging of any the isometries is equivalent up to reparametrizations.

For definiteness, let us dualize along $U(1)_R$, which corresponds to $\epsilon=0,\eta=i\lambda$ in \eqref{eqn.isom11}, with $\lambda$ a real parameter. This isometry can be gauged with a LVM (see Appendix \ref{sec.tduality}). The invariant combinations of fields are $\phi+\bar\phi$, $\chi+\bar\chi$ and $i(\bar\phi-\phi+\chi-\bar\chi)$, which are respectively gauged with the components $V^\phi$, $V^\chi$ and $V'$ of the LVM. There is a subtlety involved in dualizing from $K^{(1,1)}_0$ \eqref{eqn.kbilp1}: performing the gauging prescription of \cite{Rocek:1991ps} on $K^{(1,1)}_0$ (and gauge fixing $\phi\to0,\chi\to0$) yields 
\begin{equation}
\tilde K_0^{(1,1)} = -\frac{1}{2}(V^\chi)^2 + \int^{V^\phi-V^\chi} dq\ \log(1+e^q) - V^IX_I,
\end{equation}
in which $V'$ appears only linearly and hence cannot be solved for. To get around this issue, we restrict the potential to be defined only on the overlap of the two patches (where all the entries of the group element are nonzero), and add generalized K\"ahler transformations to the potential.\footnote{Adding these terms, which in general shifts the $b$-field of the original geometry, correspond to holomorphic symplectomorphisms of the T-dual geometry.} For simplicity, we shall consider only invariant potentials, and in that case, the generalized K\"ahler transformations which can be added must be functions of $i(\phi-\chi)$, $i(\phi+\bar\chi)$ (and complex conjugates),
\begin{equation}
K^{(1,1)} \mapsto K^{(1,1)}+f(i(\phi-\chi))+\bar f(-i(\bar\phi-\bar\chi))+g(i(\phi+\bar\chi))+\bar g(-i(\bar\phi+\chi)).
\end{equation}
These combinations are gauged by the complex gauge fields $\bV$ and $\tilde\bV$ respectively, so the T-dual potential is now
\begin{equation}
\tilde K^{(1,1)} = -\frac{1}{2}(V^\chi)^2 + f(\bV) + \bar f(\bar\bV) + g(\tilde\bV) + \bar g(\bar{\tilde\bV}) + \int^{V^\phi-V^\chi}dq\ \log(1+e^q).
\end{equation}
If we restrict to the case where $f(x)=\alpha x^2$ and $g(x)=\beta x^2$ are quadratic monomials, then solutions for all the gauge fields $V^\phi,V^\chi,V'$ exist if
\begin{equation}
8\beta\bar\beta + 2(\alpha+\bar\alpha)(\beta+\bar\beta) + (\alpha+\bar\alpha+\beta+\bar\beta) \neq 0.
\end{equation}
In order to arrive at the type $(0,0)$ potential $K^{(0,0)}_0$ obtained in \eqref{eqn.klr1}, we choose $\alpha=\tfrac{1}{2}$ and $\beta=-\tfrac{1}{4}$. The dual potential becomes
\begin{equation}
\tilde K^{(1,1)} = \frac{1}{8}(V')^2-\frac{3}{8}(V^\phi-V^\chi)^2 + \frac{1}{4}(V^\phi-V^\chi)(V^\phi+V^\chi) + \int^{V^\phi-V^\chi}dq\ \log(1+e^q) - V^IX_I.
\end{equation}
Defining the Lagrange multipliers by
\begin{equation}\begin{array}{rl}
\frac{1}{2}(X_\phi+X_\chi) &= -\frac{1}{4}(r+\bar r), \\
\frac{1}{2}(X_\phi-X_\chi) &= -\frac{1}{2}(\tell+\bar\tell), \\
X' &= \frac{i}{4}(-r+\bar r-2\tell+2\bar\tell),
\end{array}\end{equation}
after some simplification the dual potential may be written as
\begin{equation}
\tilde K^{(1,1)} = -(\tell+r)(\bar\tell+\bar r)+\int^{-r-\bar r}dq\ \log(1+e^q) + \frac{1}{2}(\tell^2+\bar\tell^2)-\frac{1}{4}(r^2+\bar r^2),
\end{equation}
which agrees with \eqref{eqn.klr1} up to a generalized K\"ahler transformation.

Performing the duality along the $U(1)$ factor requires a subtle maneuver (due to the fact that $r$ is invariant - see \eqref{eqn.isom00}) which we shall illustrate here. The $U(1)$ isometry corresponds to $\im\epsilon=0$, $\im\eta=0$ and $\epsilon+\eta=\lambda$ with $\lambda$ a real parameter. The invariant combinations $i(\phi-\bar\phi)$, $i(\chi-\bar\chi)$ and $\phi+\bar\phi-\chi-\bar\chi$ are gauged with $V^\phi$, $V^\chi$ and $V'$ respectively. Starting with \eqref{eqn.kbilp1}, we add the generalized K\"ahler transformation $\tfrac{1}{2}(\phi-\bar\chi)^2+\tfrac{1}{2}(\bar\phi-\chi)^2$, resulting in the T-dual potential
\begin{equation}
\tilde K^{(1,1)} = -\frac{1}{2}(V^\chi)^2 + \frac{1}{2}\tilde\bV^2+\frac{1}{2}\bar{\tilde\bV}^2+\int^{-\tilde\bV-\bar{\tilde\bV}}dq\ \log(1+e^q) - V^IX_I
\end{equation}
with Lagrange multipliers defined such that
\begin{equation}
V^IX_I = V^\chi i(\ell-\bar\ell) + \tilde\bV(\ell+\trx) + \bar{\tilde\bV}(\bar\ell+\bar\trx).
\end{equation}
At this point, we can eliminate $V^\chi$ using its variational equation $V^\chi=i(\bar\ell-\ell)$, while $\tilde\bV$ and $\bar{\tilde\bV}$ are somewhat more complicated functions of $(\trx+\ell)$ and $(\bar\trx+\bar\ell)$. Rather than solve them explicitly, we instead change the polarization\footnote{In superspace language, changing the polarization is a semi-chiral duality.} from $\trx,\bar\trx$ to $\rx,\bar\rx$:
\begin{align}
\tilde K'(\ell,\bar\ell,\rx,\bar\rx) &= \tilde K^{(1,1)}(\ell,\bar\ell,\trx,\bar\trx) + \trx\rx + \bar\trx\bar\rx \nn \\
&= -\frac{1}{2}(\ell-\bar\ell)^2 +\frac{1}{2}\tilde\bV^2+\frac{1}{2}\bar{\tilde\bV}^2 + \int^{-\tilde\bV-\bar{\tilde\bV}}dq\ \log(1+e^q) \nn\\&~~~~~~- \tilde\bV(\ell+\trx)-\bar{\tilde\bV}(\bar\ell+\bar\trx) + \trx\rx + \bar\trx\bar\rx
\end{align}
Since
\begin{equation}
\frac{\p\tilde K^{(1,1)}}{\p\trx} = \left(\frac{\p K^{(1,1)}}{\p V^I}-X_I\right)\frac{\p V^I}{\p\trx}-\tilde\bV,
\end{equation}
and the expression in the parentheses vanishes, the variational equation of $\trx$ sets $\rx=\tilde\bV$. This yields
\begin{equation}
\tilde K' = \ell\bar\ell-\ell\rx-\bar\ell\bar\rx+\int^{-\rx-\bar\rx}dq\ \log(1+e^q) + \frac{1}{2}(-\ell^2-\bar\ell^2+\rx^2+\bar\rx^2).
\end{equation}
A generalized K\"ahler transformation cancelling the last term on the line above, and a further change of polarization, this time on the left semi-chiral fields, brings this to \eqref{eqn.klr1}:
\begin{equation}
K^{(0,0)}_0(\tell,\bar\tell,\rx,\bar\rx) = \tilde K'(\ell,\bar\ell,\rx,\bar\rx) + \frac{1}{2}(\ell^2+\bar\ell^2-\rx^2-\bar\rx^2) - \ell\tell - \bar\ell\bar\tell.
\end{equation}

\section{$SU(2)\times SU(2)$}
\setcounter{equation}{0}
In this Appendix, we examine the two types of generalized K\"ahler structures on $SU(2)\times SU(2)$ and relate them by T-duality along an affine isometry. The complex structures on $SU(2)\times SU(2)$ were first written down in \cite{Ivanov:1994ec} and the generalized geometry is discussed in great detail in \cite{Sevrin:2011mc}.

As discussed in Section \ref{sec.22wzw}, $SU(2)\times SU(2)$ admits generalized K\"ahler structures of two types: choosing the complex structures to be equal on the Lie algebra leads to a type $(N_c,N_t)=(1,0)$ generalized K\"ahler structure while choosing them to be opposite on the Cartan subalgebra leads to a type $(0,1)$ structure.

We choose the basis $\{h,\bar h,e_1,\bar e_1,e_2,\bar e_2\}$ for the Lie algebra of $SU(2)\times SU(2)$, where
\begin{equation}
h = \tfrac{1}{2}(\sigma_{1,3}+i\sigma_{2,3}),~~~~~ e_1 = \tfrac{1}{2}(\sigma_{1,1}+i\sigma_{1,2}),~~~~~ e_2 = \tfrac{1}{2}(\sigma_{2,1}+i\sigma_{2,2})
\end{equation}
and $\sigma_{1,i}$ (respectively $\sigma_{2,i}$), $i=1,2,3$, are the sigma matrices of the first (second) $SU(2)$ factor, and the bar denotes hermitian conjugation. The two complex structures on the Lie algebra compatible with the choice of Cartan subalgebra $h,\bar h$ are
\begin{equation}
\bJ_1 = \operatorname{diag}(i,-i,i,-i,i,-i)~~~~~\text{ and }~~~~~ \bJ_2 = \operatorname{diag}(-i,i,i,-i,,-i).
\end{equation}

\subsection{Type $(1,0)$}
If one takes $J_+$ and $J_-$ both induced from the same Lie algebra complex structure, say $\bJ_1$, then generically the resulting generalized K\"ahler structure has type $(1,0)$ (there are once again positive codimension type change loci). Denoting the group element in the defining representation by $(g_{ij}^1,g^2_{ij})$, $i,j=1,2$, the chiral coordinate is
\begin{equation}
\phi = -\log g_{12}^1 + i \log g_{12}^2,
\end{equation}
while $J_\pm$ coordinates on each symplectic leaf can be chosen to be
\begin{equation}\begin{array}{ll}
z_+^1 = \ds\log\frac{g_{12}^1}{g_{22}^1}, ~~~~~~ &z_+^2 = \ds\log\frac{g_{12}^2}{g_{22}^2}, \\[5pt]
z_-^1 = \ds\log\frac{g_{12}^1}{g_{11}^1}, ~~~~~~ &z_-^2 = \ds\log\frac{g_{12}^2}{g_{11}^2}.
\end{array}
\label{eqn.su2semi}
\end{equation}
The Poisson structure in these coordinates is
\begin{equation}
\sigma(dz_\pm^1,dz_\pm^2) = \pm i.
\end{equation}
We choose the semi-chiral coordinates
\begin{equation}
\begin{array}{ll}
\ell = z_+^1, ~~~~ &\tell = -i z_+^2, \\
\rx = z_-^2, &\trx = iz_-^1,
\end{array}
\end{equation}
which satisfy $d(\tell d\ell+\bar\tell d\bar\ell+\trx d\rx+\bar\trx d\bar\rx)=0$. In the polarization determined by $\ell$ and $\rx$, the generalized potential is\footnote{Note that, due to the relations $z_+^j+\bar z_+^j=z_-^j+\bar z_-^j$ for each $j=1,2$, the combinations $(\ell,\bar\ell,\trx,\bar\trx)$ and $(\tell,\bar\tell,\rx,\bar\rx)$ are not functionally independent.}
\begin{align}
K &= F(\phi,\bar\phi) + \int^{(\ell,\bar\ell,\rx,\bar\rx)}\tell\ d\ell+\bar\tell\ d\bar\ell+\trx\ d\rx+\bar\trx\ d\bar\rx \nn \\
&=\frac{1}{2}(\bar\phi-\phi)^2 - (\rx+i\ell)(\bar\rx-i\bar\ell) -(\phi+\bar\phi)(\ell+\bar\ell)+i(\bar\phi-\phi)(\rx+\bar\rx) \nn\\&\qquad\qquad+\int^{\ell+\bar\ell}dq\ \log(1+e^q) + \int^{\rx+\bar\rx}dq\ \log(1+e^q) -\frac{1}{2}(\ell^2+\bar\ell^2+\rx^2+\bar\rx^2).
\end{align}
Here, $F(\phi,\bar\phi)=\tfrac{1}{2}(\bar\phi-\phi)^2$ is determined by solving the second order differential equations \eqref{eqn.gerbefp}-\eqref{eqn.gerbefm}. For our later discussion of T-duality, it is convenient to add generalized K\"ahler transformation terms so that
\begin{multline}
K^{(1,0)}(\ell,\bar\ell,\rx,\bar\rx,\phi,\bar\phi) = \frac{1}{2}(\bar\phi-\phi)^2 - (\rx+i\ell)(\bar\rx-i\bar\ell) + i(\bar\phi-\phi)(\rx+\bar\rx+i\ell-i\bar\ell) \\
 + \int^{\ell+\bar\ell}dq\ \log(1+e^q) + \int^{\rx+\bar\rx}dq\ \log(1+e^q) - \frac{1}{2}(\rx^2+\bar\rx^2),
\label{eqn.k10}
\end{multline}
which corresponds to shifting $\tell\to\tell'=\tell+2\phi+\ell=-iz_+^2+z_+^1+2\phi$.

\subsection{Type $(0,1)$}
If one takes $J_+$ induced from the Lie algebra complex structure $\bJ_1$ and $J_-$ induced from $\bJ_2$, then generically the resulting generalized K\"ahler structure has type $(0,1)$. The twisted chiral coordinate is
\begin{equation}
\chi = -\log g_{22}^1+i\log g_{22}^2
\end{equation}
while $J_\pm$ coordinates on the symplectic leaves can be chosen to be the same as those for the type $(1,0)$ structure \eqref{eqn.su2semi}. The Poisson structure in these coordinates, however, is now different
\begin{equation}
\sigma(dz_\pm^1,dz_\pm^2) = i.
\end{equation}
We choose now
\begin{equation}
\ell = z_+^1,~~~~~ \rx=z_-^2,~~~~~ \trx=iz_-^1,
\end{equation}
as before, but now
\begin{equation}
\tell = iz_+^2
\end{equation}
has an extra minus sign compared to the above. This is necessary to preserve $d(\tell d\ell+\bar\tell d\bar\ell+\trx d\rx+\bar\trx d\bar\rx)=0$. In the polarization determined by $\ell$ and $\rx$, the generalized potential is
\begin{align}
K &= G(\chi,\bar\chi) + \int^{(\ell,\bar\ell,\rx,\bar\rx)} \tell\ d\ell+\bar\tell\ d\bar\ell+\trx\ d\rx+\bar\trx\ d\bar\rx \nn\\
&= -\frac{1}{2}(\bar\chi-\chi)^2 + i(\ell\rx-\bar\ell\bar\rx) - (\chi+\bar\chi)(\ell+\bar\ell) + i(\bar\chi-\chi)(\rx+\bar\rx) \nn\\
&\qquad\qquad+\int^{\ell+\bar\ell}dq\ \log(1+e^q) + \int^{\rx+\bar\rx}dq\ \log(1+e^q),
\end{align}
where $G(\chi,\bar\chi)=-\tfrac{1}{2}(\bar\chi-\chi)^2$ is determined by \eqref{eqn.gerbefp}, \eqref{eqn.gerbefm}.

For later discussion of T-duality, it is convenient to add generalized K\"ahler transformation terms so that
\begin{multline}
K^{(0,1)}(\ell,\bar\ell,\rx,\bar\rx,\chi,\bar\chi) = -\frac{1}{2}(\bar\chi-\chi)^2 + i(\bar\chi-\chi)(\rx+\bar\rx+i\ell-i\bar\ell) \\
+\int^{\ell+\bar\ell}dq\ \log(1+e^q) + \int^{\rx+\bar\rx}dq\ \log(1+e^q) - \frac{1}{2}(\ell^2+\bar\ell^2)
\label{eqn.k01}
\end{multline}
which corresponds to shifting $\tell\to\tell'=\tell+2\chi-\ell=iz_+^2-z_+^1+2\chi$. This matches with $\tell$ in \eqref{eqn.k10} since $iz_+^2-z_+^1+2\chi=-iz_+^2+z_+^1+2\phi$.

\subsection{T-duality}
The subgroup of the isometry group preserving both complex structures is $(U(1)\times U(1))_L\times(U(1)\times U(1))_R$, acting as $g\mapsto e^{\epsilon\bar h-\bar\epsilon h}ge^{\bar\eta\bar h-\eta h}$. Under this action, the coordinates transform as
\begin{equation}
\begin{array}{rlrl}
\phi&\mapsto\phi-i\bar\epsilon+i\eta,\qquad\qquad& \chi&\mapsto\chi+i\bar\epsilon+i\eta, \\
z_+^1&\mapsto z_+^1+i\epsilon+i\bar\epsilon & z_-^1&\mapsto z_-^1 -i\eta-i\bar\eta, \\
z_+^2&\mapsto z_+^2+\epsilon-\bar\epsilon & z_-^2&\mapsto z_-^2+\bar\eta-\eta.
\end{array}
\end{equation}
For the type $(1,0)$ structure, the parameters satisfy
\begin{equation}
\bar {\bD}_+\epsilon=0,~~~ {\bD}_\pm\epsilon=0,~~~ \bar {\bD}_\pm\eta=0,~~~ {\bD}_-\eta=0,
\end{equation}
while for the type $(0,1)$ structure, they satisfy
\begin{equation}
\bar {\bD}_\pm\epsilon=0,~~~ {\bD}_+\epsilon=0,~~~ \bar {\bD}_\pm\eta=0,~~~ {\bD}_-\eta=0.
\end{equation}
In both cases, it is clear that these isometries are affine, $\p_\doubleplus\epsilon=0=\p_=\eta$.

Let us perform T-duality along the isometry with parameter $\epsilon=-i\lambda,\eta=0$, with $\lambda$ real. This isometry transforms $\phi\mapsto\phi+\lambda$ and $\chi\mapsto\chi-\lambda$ and leaves $\ell,\bar\ell,\rx,\bar\rx$ invariant. The potential \eqref{eqn.k10} is invariant under this isometry, and can be gauged by a standard vector multiplet. Constraining the gauge field to be flat using a twisted chiral Lagrange multiplier $\chi$ returns one to the original model
\begin{multline}
\tilde K^{(1,0)} = -\frac{1}{2}V^2 - (\rx+i\ell)(\bar\rx-i\bar\ell) + V(\rx+\bar\rx+i\ell-i\bar\ell) \\
+\int^{\ell+\bar\ell}dq\ \log(1+e^q) + \int^{\rx+\bar\rx}dq\ \log(1+e^q) - \frac{1}{2}(\rx^2+\bar\rx^2) - i(\bar\chi-\chi)V,
\end{multline}
where we have gauge fixed $\phi=0$. It is now straightforward to check that integrating out the gauge field $V$ yields the type $(0,1)$ potential $K^{(0,1)}$ \eqref{eqn.k01}.

\bibliographystyle{jhep}
\bibliography{wzw1}

\begin{thebibliography}{10}

\bibitem{Zumino:1979et}
B.~Zumino.
\newblock {Supersymmetry and Kahler Manifolds}.
\newblock {\em Phys. Lett.}, B87:203, 1979.

\bibitem{Gates:1984nk}
S.~J. Gates, Jr., C.~M. Hull, and M.~{Ro\v cek}.
\newblock {Twisted Multiplets and New Supersymmetric Nonlinear Sigma Models}.
\newblock {\em Nucl. Phys.}, B248:157--186, 1984.

\bibitem{Buscher:1987uw}
T.~Buscher, U.~{Lindstr\"om}, and M.~{Ro\v cek}.
\newblock {New Supersymmetric $\sigma$ Models With {Wess-Zumino} Terms}.
\newblock {\em Phys. Lett.}, B202:94--98, 1988.

\bibitem{Sevrin:1996jr}
Alexander Sevrin and Jan Troost.
\newblock {Off-shell formulation of N=2 nonlinear sigma models}.
\newblock {\em Nucl. Phys.}, B492:623--646, 1997.

\bibitem{Grisaru:1997ep}
Marcus~T. Grisaru, M.~Massar, A.~Sevrin, and J.~Troost.
\newblock {Some aspects of N=(2,2), D = 2 supersymmetry}.
\newblock {\em Fortsch. Phys.}, 47:301--307, 1999.

\bibitem{Lindstrom:2005zr}
Ulf {Lindstr\"om}, Martin {Ro\v cek}, Rikard von Unge, and Maxim Zabzine.
\newblock {Generalized Kahler manifolds and off-shell supersymmetry}.
\newblock {\em Commun. Math. Phys.}, 269:833--849, 2007.

\bibitem{Hitchin:2004ut}
Nigel Hitchin.
\newblock {Generalized Calabi-Yau manifolds}.
\newblock {\em Quart. J. Math.}, 54:281--308, 2003.

\bibitem{Gualtieri:2003dx}
Marco Gualtieri.
\newblock {\em {Generalized complex geometry}}.
\newblock PhD thesis, Oxford U., 2003.

\bibitem{Grana:2004bg}
Mariana Grana, Ruben Minasian, Michela Petrini, and Alessandro Tomasiello.
\newblock {Supersymmetric backgrounds from generalized Calabi-Yau manifolds}.
\newblock {\em JHEP}, 08:046, 2004.

\bibitem{Jeschek:2004wy}
Claus Jeschek and Frederik Witt.
\newblock {Generalised G(2) - structures and type IIb superstrings}.
\newblock {\em JHEP}, 03:053, 2005.

\bibitem{Buscher:1987qj}
T.~H. Buscher.
\newblock {Path Integral Derivation of Quantum Duality in Nonlinear Sigma
  Models}.
\newblock {\em Phys. Lett.}, B201:466--472, 1988.

\bibitem{Ivanov:1994ec}
Ivan~T. Ivanov, Byung-bae Kim, and Martin {Ro\v cek}.
\newblock {Complex structures, duality and WZW models in extended superspace}.
\newblock {\em Phys. Lett.}, B343:133--143, 1995.

\bibitem{Rocek:1991ps}
Martin {Ro\v cek} and Erik~P. Verlinde.
\newblock {Duality, quotients, and currents}.
\newblock {\em Nucl. Phys.}, B373:630--646, 1992.

\bibitem{Spindel:1988sr}
P.~Spindel, A.~Sevrin, W.~Troost, and Antoine Van~Proeyen.
\newblock {Extended Supersymmetric Sigma Models on Group Manifolds. 1. The
  Complex Structures}.
\newblock {\em Nucl. Phys.}, B308:662--698, 1988.

\bibitem{Rocek:1991vk}
M.~{Ro\v cek}, K.~Schoutens, and A.~Sevrin.
\newblock {Off-shell WZW models in extended superspace}.
\newblock {\em Phys. Lett.}, B265:303--306, 1991.

\bibitem{Yano}
Kentaro Yano.
\newblock {\em {Differential geometry on complex and almost complex spaces}}.
\newblock Pergamon Press, 1965.

\bibitem{Hull:1985pq}
C.~M. Hull, A.~Karlhede, U.~{Lindstr\"om}, and M.~{Ro\v cek}.
\newblock {Nonlinear $\sigma$ Models and Their Gauging in and Out of
  Superspace}.
\newblock {\em Nucl. Phys.}, B266:1--44, 1986.

\bibitem{Sevrin:2011mc}
Alexander Sevrin, Wieland Staessens, and Dimitri Terryn.
\newblock {The Generalized Kahler geometry of N=(2,2) WZW-models}.
\newblock {\em JHEP}, 12:079, 2011.

\bibitem{Lindstrom:2014bra}
Ulf Lindström.
\newblock {Extended supersymmetry of semichiral sigma models in 4D}.
\newblock {\em JHEP}, 02:170, 2015.

\bibitem{Hull:2008vw}
Chris~M. Hull, Ulf {Lindstr\"om}, Martin {Ro\v cek}, Rikard von Unge, and Maxim Zabzine.
\newblock {Generalized Kahler geometry and gerbes}.
\newblock {\em JHEP}, 10:062, 2009.

\bibitem{Grisaru:1997pg}
Marcus~T. Grisaru, M.~Massar, A.~Sevrin, and J.~Troost.
\newblock {The Quantum geometry of N=(2,2) nonlinear sigma models}.
\newblock {\em Phys. Lett.}, B412:53--58, 1997.

\bibitem{Hull:2010sn}
Chris~M. Hull, Ulf {Lindstr\"om}, Martin {Ro\v cek}, Rikard von Unge, and Maxim Zabzine.
\newblock {Generalized Calabi-Yau metric and Generalized Monge-Ampere
  equation}.
\newblock {\em JHEP}, 08:060, 2010.

\bibitem{Lindstrom:2007vc}
Ulf {Lindstr\"om}, Martin {Ro\v cek}, Itai Ryb, Rikard von Unge, and Maxim Zabzine.
\newblock {New N = (2,2) vector multiplets}.
\newblock {\em JHEP}, 08:008, 2007.

\bibitem{Lindstrom:2007sq}
Ulf {Lindstr\"om}, Martin {Ro\v cek}, Itai Ryb, Rikard von Unge, and Maxim Zabzine.
\newblock {T-duality and Generalized Kahler Geometry}.
\newblock {\em JHEP}, 02:056, 2008.

\bibitem{Lindstrom:2008hx}
Ulf {Lindstr\"om}, Martin {Ro\v cek}, Itai Ryb, Rikard von Unge, and Maxim Zabzine.
\newblock {Nonabelian Generalized Gauge Multiplets}.
\newblock {\em JHEP}, 02:020, 2009.

\end{thebibliography}

\end{document}